\documentclass[lettersize,journal]{IEEEtran}
\usepackage{array}
\newcolumntype{L}[1]{>{\raggedright\let\newline\\\arraybackslash\hspace{0pt}}m{#1}}
\usepackage{textcomp}
\usepackage{stfloats}
\usepackage{verbatim}
\usepackage{cite}
\usepackage{xcolor}

\usepackage[colorlinks=true,linkcolor=black,citecolor=black,urlcolor=blue]{hyperref}

\usepackage{microtype}
\usepackage{graphicx}
\usepackage{subfigure}
\usepackage{booktabs} 
\usepackage{bbm}

\usepackage{pgfplots}

\usepackage{amsmath,amssymb,amsfonts}
\usepackage{xurl}
\usepackage{stackengine}
\usepackage{tikz}
\usetikzlibrary{decorations.pathreplacing}
\usetikzlibrary{positioning,arrows.meta,quotes}
\usetikzlibrary{shapes,snakes}
\usetikzlibrary{bayesnet}
\tikzset{>=latex}
\tikzstyle{plate caption} = [caption, node distance=0, inner sep=0pt, below left=5pt and 0pt of #1.south]

\usepackage[normalem]{ulem}
\usepackage{multirow}
\usepackage{mathtools}

\title{Bayesian Calibration of the Intelligent Driver Model}

\author{Chengyuan Zhang,~\IEEEmembership{Graduate Student Member,~IEEE,}
        and~Lijun Sun,~\IEEEmembership{Senior Member,~IEEE}
\thanks{(Corresponding author: Lijun Sun)}
\thanks{C. Zhang and L. Sun are with the Department of Civil Engineering, McGill University, Montreal, QC H3A 0C3, Canada. E-mail:
enzozcy@gmail.com (C. Zhang), lijun.sun@mcgill.ca (L.
Sun).}
\thanks{The authors would like to thank the McGill Engineering Doctoral Awards (MEDA), the Institute for Data Valorisation (IVADO), the Interuniversity Research Centre on Enterprise Networks, Logistics and Transportation (CIRRELT), Fonds de recherche du Québec -- Nature et technologies (FRQNT), and the Natural Sciences and Engineering Research Council (NSERC) of Canada for providing scholarships and funding to support this study.}
}

\begin{document}
\maketitle

\begin{abstract}
Accurate calibration of car-following models is essential for understanding human driving behaviors and implementing high-fidelity microscopic simulations. This work proposes a memory-augmented Bayesian calibration technique to capture both uncertainty in the model parameters and the temporally correlated behavior discrepancy between model predictions and observed data. Specifically, we characterize the parameter uncertainty using a hierarchical Bayesian framework and model the temporally correlated errors using Gaussian processes. We apply the Bayesian calibration technique to the intelligent driver model (IDM) and develop a novel stochastic car-following model named memory-augmented IDM (MA-IDM). To evaluate the effectiveness of MA-IDM, we compare the proposed MA-IDM with Bayesian IDM in which errors are assumed to be i.i.d., and our simulation results based on the HighD dataset show that MA-IDM can generate more realistic driving behaviors and provide better uncertainty quantification than Bayesian IDM. By analyzing the lengthscale parameter of the Gaussian process, we also show that taking the driving actions from the past five seconds into account can be helpful in modeling and simulating the human driver’s car-following behaviors.
\end{abstract}


\section{Introduction}
Microscopic car-following models are powerful tools to study and simulate human driving behaviors in traffic flows at the trajectory level. They reveal the mechanisms of complex interactions between a subject vehicle and its leading vehicle. These interactions are the essential factors that affect the dynamics of traffic flow and create diverse macroscopic traffic phenomena \cite{wang2022social}. 
Although experiences indicate that we cannot calibrate a zero error model that perfectly fits the data \cite{punzo2021calibration}, some car-following models are still valid with which one can adopt various calibration (or parameter identification) methods to reproduce reality to different extents. Given that the performance and fidelity of microscopic car-following models are heavily dependent on accurate calibration, improving the quality of calibration methods is a critical research question.

In recent studies, probabilistic calibration methods have become an emerging and promising approach with a solid statistical foundation. There are essentially two probabilistic approaches for model calibration. The first is maximum likelihood estimation (MLE): we often solve MLE as an optimization problem and finally obtain a point estimation for model parameters; see, e.g., \cite{hoogendoorn2010calibration}. The second is Bayesian inference, which allows us to exploit the full posterior distribution for model parameters through Markov chain Monte Carlo (MCMC) or variational inference (VI); see, e.g., \cite{rahman2015improving,abodo2019strengthening}. In general, the Bayesian inference approach is advantageous in two aspects: (1) we can perform uncertainty quantification based on the full posterior distribution. This is particularly important for a simulation model as we are often interested in the distribution and uncertainty of the simulation results; (2) Bayesian inference offers a hierarchical modeling scheme that allows us to learn parameter distributions at both population and individual levels, where the population distribution prevents overfitting by imposing certain dependencies on the parameters. This is particularly important for calibrating car-following models since samples are often short trajectories collected from a large number of drivers with diverse vehicle configurations/dynamics. Considering substantial differences in personal driving styles and vehicle configurations/dynamics, it is critical to learn the population distribution that can generalize the parameter distribution from short trajectory data. In addition, the learned population posterior distribution can help create diverse driving behaviors to create more realistic simulations with driver heterogeneity. Note that in the following of this paper, we refer to ``driver behavior'' as the characteristics resulting from both the driver and the vehicle.

\begin{figure}[t]
    \centering
    \centering
    \includegraphics[width = 0.85\linewidth]{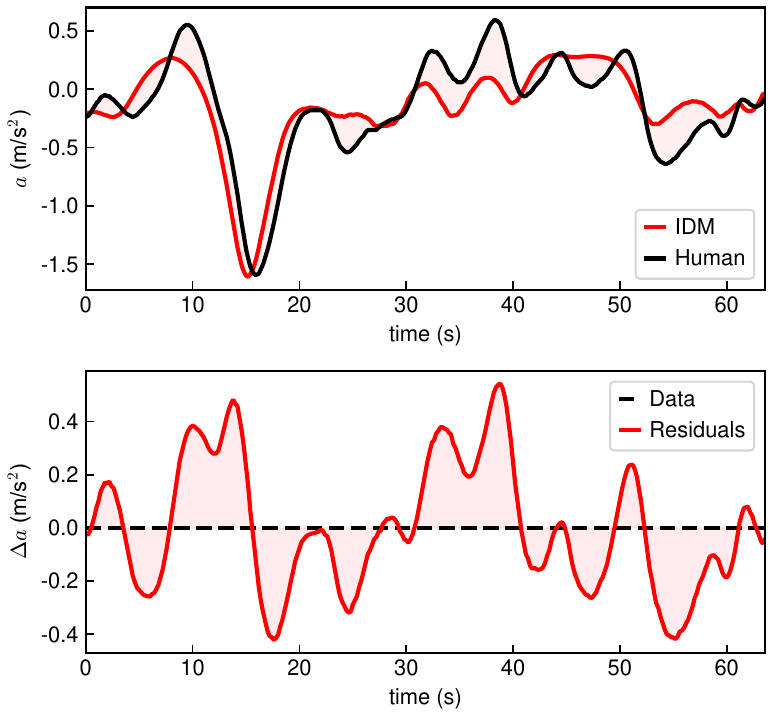}
    \caption{The first panel shows the predicted acceleration from a well-calibrated IDM and the real-world observation from a human driver. The second panel shows the corresponding discrepancy. Note that the literature usually assumes $a_\text{Human}=a_\text{IDM}+\text{error}$, where the error is assumed \textit{i.i.d.} However, we can clearly see strong serial correlations in the residual. Ignoring such serial correlations will undoubtedly lead to biased modeling fitting.}
    \label{noniid}
\end{figure}

A fundamental question in performing probabilistic calibration is how to define the probabilistic model and data generation process. Most existing car-following models seek parsimonious structures by simply taking observations from the most recent (i.e., only one) step as input to generate acceleration/speed as output for the current step. However, given physical inertia, delay in reaction, and missing important covariates (e.g., car follower models, in general, do not take the status of the breaking light of the leading vehicle as input variable), we should expect unexplained behavior (ie, the discrepancy between predicted acceleration/speed and observed acceleration/speed) to be temporally correlated \cite{morton2016analysis,wang2017capturing}. For instance, Wang et al. \cite{wang2017capturing} show that the best-performed deep learning model essentially takes states/observations from the most recent
$\sim$10 sec as input. However, most existing probabilistic calibration methods are developed based on a simple assumption---the errors are independent and identically distributed (\textit{i.i.d.}), and as a result, ignoring the autocorrelation in the residuals will lead to biased calibration \cite{treiber2013microscopic,punzo2020two}. For example, Fig.~\ref{noniid} shows the residual process in acceleration ($a$), speed ($v$), and gap ($s$) when calibrating an IDM model with \textit{i.i.d.} noise, and we can see that the residuals have strong serial correlation (i.e., autocorrelation). Essentially, two classes of methods are developed in the literature to model serial correlation. One approach is to directly process the time series data to eliminate serial correlations. For instance, Hoogendoorn and Hoogendoorn \cite{hoogendoorn2010calibration} adopted a difference transformation (see \cite{cochrane1949application}) to eliminate the serial correlation. However, they used empirical correlation coefficients to perform the transformation instead of jointly learning the IDM parameters and the correlations. Another approach is to explicitly model the serial correlations (e.g., by stochastic processes); for example, Treiber and Kesting \cite{treiber2013traffic} introduced the Wiener process to model the temporally correlated error process.

In this paper, we propose a novel probabilistic calibration approach that takes advantage of Bayesian methods by capturing both uncertainty in the model parameters and the temporally correlated behavior discrepancy. We apply this approach to calibrate the IDM and develop an MA-IDM, which models the serially correlated errors as zero-mean Gaussian processes (GP). Taking advantage of the Bayesian methods, we can jointly learn the model parameters and the GP hyperparameters by MCMC. In this work, the MA-IDM is presented in three forms, i.e., pooled model, hierarchical model, and unpooled model. We conducted numerical experiments on the Highway Drone (HighD) data set \cite{highDdataset}, and the results demonstrate that our hierarchical model not only learns consistent driving styles at the population level but also depicts heterogeneity of driving behavior at the individual level. Additionally, a stochastic simulation method is developed to obtain the posterior motion states in terms of acceleration, speed, and location.

The overall contributions of this work are threefold:
\begin{enumerate}
    \item We develop a novel Bayesian calibration approach to learn unbiased parameters and their full posterior distribution, in which GP is introduced to model autocorrelated errors. This approach is applied to calibrate IDM. The autocorrelation in the residuals reveals that incorporating observations from the past $\sim 5$ seconds improves the modeling of car-following decisions.
    \item With a hierarchical MA-IDM, heterogeneous but consistent driving behaviors can be learned at the individual level. Therefore, we can generate enormous drivers with heterogeneous driving behaviors/styles governed by the same population distribution. In such a way, our model can help create simulations with driver/car heterogeneity.
    \item We introduce an unbiased stochastic simulator, which is inspired by the corresponding generative process of our Bayesian calibration approach. As a result, the simulator can produce more realistic results and better uncertainty quantification than those with homogeneous parameters or random parameters.
\end{enumerate}

The structure of the remaining contents is organized as follows. First, Section~\ref{related_works} introduces related literature on calibrating car-following models and modeling serial correlations. Section~\ref{Preliminaries} emphasizes some preliminaries and formulates the car-following problem based on IDM. Then in Section~\ref{method}, we propose a novel Bayesian calibration method and develop several novel car-following models, i.e., the Bayesian-IDM (B-IDM) and the MA-IDM, with different hierarchies. Next, we conduct extensive experiments and simulations, then thoroughly analyze the results in Section~\ref{experiments} and Section~\ref{simulation_sec}. Finally, Section~\ref{conclusion} wraps up this work and discusses several potential directions worth further exploration.

\section{Related Work}\label{related_works}
\subsection{Car-Following Model Calibration}
Proper choice of key parameters in a car-following model can help to depict and reproduce complex driving behaviors. As in most cases, we will never know the values of these parameters. Typically, one could collect observable measurements from the field car-following data in diverse scenarios and fit the model to the observed data by adjusting the parameters to optimize certain objective functions. The overall process is known as model calibration \cite{kennedy2001bayesian}. We refer readers to \cite{punzo2021calibration} for a review of the calibration of car-following models.

Various calibration methods have been studied in the literature. The genetic algorithm (GA) is one of the most typical and traditional tools as a heuristic evolutionary algorithm \cite{punzo2021calibration, chen2010calibration}. For example, Punzo et al. \cite{punzo2021calibration} conducted extensive experiments to study $29$ goodness-of-fit functions (GoF) with combinations of three measures of performance (MoP) and provided systematic guidelines on GA-based calibration. The combinations of GoF could be selected according to the specific problem and traffic scenarios, leading to the multi-objective calibration approach \cite{he2022physics, zheng2023multi}. However, as an optimization tool, GA presents some limitations because it is not only computationally expensive but also sensitive to different choices/combinations of GoF and MoP. Besides, the best combination of one dataset may not be suitable for another. Maximum likelihood (MLE) is a powerful approach. Hoogendoorn et al. \cite{hoogendoorn2005parameter, hoogendoorn2010calibration} first proposed to calibrate car-following models based on MLE. Treiber and Kesting \cite{treiber2013microscopic} thoroughly investigated the MLE approach on the calibration and validation of car-following models. MLE seeks to estimate a single ``best'' value of an unknown quantity based on optimization algorithms. The Bayesian method, as an alternative approach, aims to estimate the distribution of an unknown variable (i.e., posterior). Rahman et al. \cite{rahman2015improving} calibrated car-following models based on a Bayesian approach. The experiments showed that the Bayesian approach provided much better results than the deterministic optimization algorithm. With the Bayesian approach, Abodo et al. \cite{abodo2019strengthening} further developed a hierarchical model to calibrate the IDM, which has also achieved promising results.

\subsection{Serial Correlation Modeling}
Regression plays a vital role in model calibration. In the literature, many regression models are developed heavily based on an assumption---the errors are \textit{i.i.d.} random variables. However, due to the omission of relative covariates and model misspecification, we often see autocorrelated errors when estimating regression models on time series data. In this case, assuming the errors to be  \textit{i.i.d.} will lead to suboptimal parameter identification. Generally, as mentioned in \cite{kennedy2001bayesian}, the observations are usually regarded as the {true process} with an \textit{i.i.d.} {observation error}; while in calibration methods, observations are modeled as the combination of a {calibrated model} and an independent {model inadequacy function} with an \textit{i.i.d.} {observation error}. In calibrations involving time series data, the \textit{model inadequacy function} is mainly used to capture the autocorrelations.

In general, there are two ways to perform model estimation in the presence of serial correlations: (1) by directly processing the nonstationary data and eliminating serial correlations (e.g., performing the differencing operation), so that one can safely ignore the \textit{model inadequacy function} and obtain stationary time series;  or (2) by explicitly modeling the serial correlations based on specific \textit{model inadequacy functions}. For instance, Hoogendoorn \cite{hoogendoorn2010calibration} performed a differencing transformation to eliminate serial correlations, which then did not show significant differences between autocorrelation coefficients and zeros in the previously mentioned Durbin–Watson test \cite{durbin1950testing}. However, the information conveyed by the serial correlations is directly discarded, which prevents us from modeling the generative processes of observations. Another way is to explicitly develop the formation of serial correlations based on further assumptions. For example, dynamic regression models leverage linear regression and autoregressive integrated moving average (ARIMA) into a single regression model to forecast time series data \cite{hyndman2018forecasting}. For the main scope of this paper, i.e., calibration and simulation of car-following models, we use GP \cite{rasmussen2003gaussian} to model serially correlated errors (i.e., for the \textit{model inadequacy} part). GP provide a solid statistical solution to learn the autocorrelation structure, and more importantly, it allows us to understand the temporal effect in driving behavior through the lengthscale parameter $l$, which partially explains the memory effect (see \cite{treiber2003memory}) of human driving behaviors.

\begin{figure}[t]
    \centering
    \includegraphics[width = \linewidth]{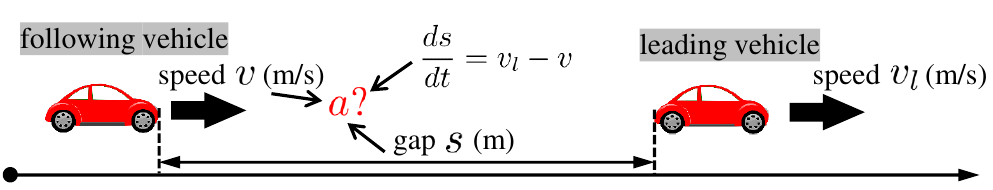}
    \caption{Physical settings of a car-following scenario.}
    \label{cf_illustration}
\end{figure}

\section{Preliminaries and Problem Formulation}\label{Preliminaries}
\subsection{IDM and Probabilistic IDM}
The traditional IDM \cite{treiber2000congested} is a continuous nonlinear function $f: \mathbb{R}^3 \mapsto \mathbb{R}$ which maps the gap, the speed, and the speed difference (approach rate) to acceleration $a_{\mathrm{IDM}}$ at a certain time point. Here, we denote $s$ as the gap between the following vehicle and the leading vehicle, $v$ as the speed of the following vehicle, and $\Delta v=-ds/dt=v-v_l$ ($v_l$ denotes the speed of the leading vehicle) as the speed difference. The physical meaning of these notation is illustrated in Fig.~\ref{cf_illustration}. With these notations, IDM is defined as
\begin{align}\label{IDM_eq}
\begin{split}
    & a_{\mathrm{IDM}} = f(s,v,\Delta v) \triangleq \alpha\,\left(1-\left({\frac{v}{v_{0}}}\right)^{\delta }-\left({\frac{s^{\ast}(v,\Delta v)}{s}}\right)^{2}\right),
\end{split}\\
\begin{split}\label{IDM_eq_gap}
  s^{\ast}(v,\Delta v) =s_{0}+ s_1\,\sqrt{\frac{v}{v_0}} + v\,T+{\frac{v\,\Delta v}{2\,{\sqrt{\alpha\,\beta}}}},
\end{split}
\end{align}
where, $v_0,  s_{0}, T,  \alpha, \beta,$ and $\delta$ are model parameters with the following meaning: the desired speed $v_{0}$ is the free-flow speed; the jam spacing $s_{0}$ denotes a minimum gap distance from the leading vehicle; the safe time headway $T$ represents the minimum time interval between the following vehicle and the leading vehicle; the acceleration $\alpha$ and the comfortable braking deceleration $\beta$ are the maximum vehicle acceleration and the desired deceleration to keep safe, respectively; the exponential coefficient $\delta$ is a constant, usually set as $4$ at default \cite{treiber2000congested}. In IDM, the deceleration is controlled by the desired minimum gap $s^\ast$ in Eq.~\eqref{IDM_eq_gap}, in which we set $s_1=0$ following \cite{treiber2000congested} to obtain a model with interpretable parameters.

To make the notation concise and compact, we denote the IDM parameters by $\boldsymbol{\theta}=[v_0,s_0,T,\alpha,\beta] \in \mathbb{R}^5$. For a certain vehicle $d$, we have the IDM actions $a_{\mathrm{IDM},d}^{(t)} = f(s_d^{(t)},v_d^{(t)},\Delta v_d^{(t)};\boldsymbol{\theta}_d)$, where the subscript $d$ represents the index for each driver, and the superscript $(t)$ indicates the timestamp. Compactly, we denote the inputs at time $t$ as a vector $\boldsymbol{h}_d^{(t)}=[s_d^{(t)},v_d^{(t)},\Delta v_d^{(t)}], \forall t\in\{t_0, \dots, t_0+(T-1)\Delta t\}$. Here, we adopt the scheme in \cite{treiber2013microscopic, treiber2006delays} with a step of $\Delta t$ to update vehicle speed and location as following:
\begin{subequations}\label{simulation_update}
    \begin{align}
        v^{(t+\Delta t)} &= v^{(t)} + a^{(t)} \Delta t,\label{update_v}\\
        x^{(t+\Delta t)} &= x^{(t)} + v^{(t)} \Delta t + \frac{1}{2} a^{(t)} \Delta t^2.\label{update_x}
    \end{align}
\end{subequations}
However, real-world driving actions cannot be fully modeled by IDM and thus it is inevitable to observe some discrepancies, as stated by Treiber and Kesting (see Section 3.2 of Chapter 12 in \cite{treiber2013traffic}). It can be modeled by adding some acceleration noise of standard deviation $\sigma_{\epsilon}$ to $a_{\text{IDM}}$. In such a setting, we consider $a_{\text{IDM}}$ as a rational behavior model, while the random term as the imperfect driving behaviors that cannot be modeled by IDM. By taking the random term with \textit{i.i.d} assumption into consideration, one can develop a stochastic version of IDM, i.e., the probabilistic IDM \cite{bhattacharyya2020online, treiber2017intelligent}, given by
\begin{equation}\label{iid_a}a_d^{(t)}|\boldsymbol{h}_d^{(t)},\boldsymbol{\theta}_d\, \stackon{$\sim$}{\tiny{\textit{i.i.d.}}}\, \mathcal{N}(a_{\mathrm{IDM},d}^{(t)},\sigma_{\epsilon}^2),
\end{equation}
where $a_d^{(t)}$ is the true acceleration, and $\mathcal{N}(\mu,\sigma^2)$ denotes a Gaussian distribution with $\mu$ as the mean and $\sigma^2$ as the variance.

The calibration of the IDM mainly concerns the identification of the model parameters $\boldsymbol{\theta}$ to learn the best mapping from $\boldsymbol{h}_d^{(t)}$ to specific observations \cite{punzo2021calibration}, e.g., ${x}_d^{(t+\Delta t)}$ and ${v}_d^{(t+\Delta t)}$. 

\begin{figure*}[!t]
    \centering
    \subfigure{\centering
    \resizebox{0.9\linewidth}{!}{\tikzset{every picture/.style={line width=0.75pt}} 

\begin{tikzpicture}[x=0.75pt,y=0.75pt,yscale=-1,xscale=1]

\draw [dotted, thin, color=blue] (-70,25) -- (560,25);

\draw [dotted, thin, color=red] (-70,-25) -- (560,-25);

\node [text width=1.8cm, color=black] () at (520,50) {{observation}};

\node [text width=3.5cm, color=blue] () at (540,0) {{individual level\\(driving behaviors)}};

\node [text width=3.5cm, color=red] () at (545,-50) {{population level\\(driving styles)}};

\node [circle,draw=red,fill=white, inner sep=0pt,minimum size=0.75cm] (o_pool1) at (0,-50) {$\boldsymbol{\theta}$};
\node [text width=4cm, color=black] () at (20,80) {(a) Pooled model.};

\node [obs, draw=black,inner sep=0pt,minimum size=0.75cm] (y_pool1) at (-60,50) {$y_1$};

\node [obs, draw=black,inner sep=0pt,minimum size=0.75cm] (y_pool2) at (-20,50) {$y_2$};

\node [text width=0.5cm, color=black] () at (20,50) {$\cdots$};

\node [obs, draw=black,inner sep=0pt,minimum size=0.75cm] (y_poolD) at (60,50) {$y_D$};

\path [draw,->] (o_pool1) edge (y_pool1);
\path [draw,->] (o_pool1) edge (y_pool2);
\path [draw,->] (o_pool1) edge (y_poolD);

\draw [->, very thick] (125,-45)  -- (55,-45);
\node [color=black] () at (90,-59) {\color{blue} $\mathrm{Var}(\boldsymbol{\theta}_d)$: small variance};

\draw [->, very thick] (250,-45)  -- (320,-45);
\node [color=black] () at (285,-59) {\color{blue} $\mathrm{Var}(\boldsymbol{\theta}_d)$: large variance};

\node [circle,draw=red,fill=white, inner sep=0pt,minimum size=0.75cm] (o_par_pool_all) at (190,-50) {$\boldsymbol{\theta}$};
\node [text width=4.5cm, color=black] () at (210,80) {(b) Hierarchical model.};

\node [circle,draw=blue,fill=white, inner sep=0pt,minimum size=0.75cm] (o_par_pool1) at (130,0) {$\boldsymbol{\theta}_1$};

\node [obs, draw=black,inner sep=0pt,minimum size=0.75cm] (y_par_pool1) at (130,50) {$y_1$};

\node [circle,draw=blue,fill=white, inner sep=0pt,minimum size=0.75cm] (o_par_pool2) at (170,0) {$\boldsymbol{\theta}_2$};

\node [obs, draw=black,inner sep=0pt,minimum size=0.75cm] (y_par_pool2) at (170,50) {$y_2$};

\node [text width=0.5cm, color=black] () at (210,0) {$\cdots$};

\node [text width=0.5cm, color=black] () at (210,50) {$\cdots$};

\node [circle,draw=blue,fill=white, inner sep=0pt,minimum size=0.75cm] (o_par_poolD) at (250,0) {$\boldsymbol{\theta}_D$};

\node [obs, draw=black,inner sep=0pt,minimum size=0.75cm] (y_par_poolD) at (250,50) {$y_D$};

\path [draw,->] (o_par_pool1) edge (y_par_pool1);
\path [draw,->] (o_par_pool2) edge (y_par_pool2);
\path [draw,->] (o_par_poolD) edge (y_par_poolD);
\path [draw,->] (o_par_pool_all) edge (o_par_pool1);
\path [draw,->] (o_par_pool_all) edge (o_par_pool2);
\path [draw,->] (o_par_pool_all) edge (o_par_poolD);

\node [circle,draw=blue,fill=white, inner sep=0pt,minimum size=0.75cm] (o_no_pool1) at (320,0) {$\boldsymbol{\theta}_1$};
\node [text width=4cm, color=black] () at (390,80) {(c) Unpooled model.};

\node [obs, draw=black,inner sep=0pt,minimum size=0.75cm] (y_no_pool1) at (320,50) {$y_1$};

\node [circle,draw=blue,fill=white, inner sep=0pt,minimum size=0.75cm] (o_no_pool2) at (360,0) {$\boldsymbol{\theta}_2$};

\node [obs, draw=black,inner sep=0pt,minimum size=0.75cm] (y_no_pool2) at (360,50) {$y_2$};

\node [text width=0.5cm, color=black] () at (400,0) {$\cdots$};

\node [text width=0.5cm, color=black] () at (400,50) {$\cdots$};

\node [circle,draw=blue,fill=white, inner sep=0pt,minimum size=0.75cm] (o_no_poolD) at (440,0) {$\boldsymbol{\theta}_D$};

\node [obs, draw=black,inner sep=0pt,minimum size=0.75cm] (y_no_poolD) at (440,50) {$y_D$};

\path [draw,->] (o_no_pool1) edge (y_no_pool1);
\path [draw,->] (o_no_pool2) edge (y_no_pool2);
\path [draw,->] (o_no_poolD) edge (y_no_poolD);

\end{tikzpicture}}}
    \caption{The pooled, hierarchical, and unpooled model. Note that the pooled model can be regarded as a model with identical $\boldsymbol{\theta}_d$, while the unpooled model is with independent $\boldsymbol{\theta}_d$.}\label{pooling_techinique}
\end{figure*}
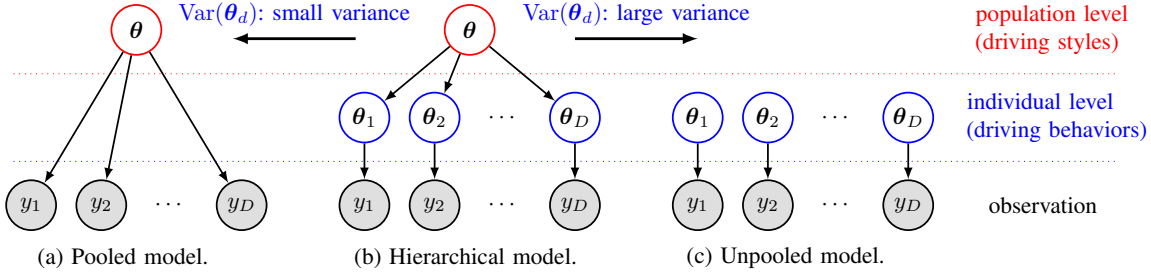

\subsection{Pooled, Hierarchical, and Unpooled Model}
Here we introduce three different hierarchies of the model, i.e., pooled, hierarchical, and unpooled models, in which hierarchical models are often considered in terms of ``partial pooling," compromising between the pooled (``complete pooling'') and unpooled (``no pooling'') models \cite{gelman2006multilevel}. Figure ~\ref{pooling_techinique} illustrates the pooling technique, highlighting differences among the three models.  In a pooled model, all drivers, denoted as $\forall d \in {1,\dots,D}$, are assumed to exhibit identical driving behaviors, characterized by a single parameter set $\boldsymbol{\theta}$. In contrast, an unpooled model assumes each driver $d$ has a distinct, unrelated parameter set $\boldsymbol{\theta}_d$.
Hierarchical models combine these approaches: each driver has unique parameters, yet all are influenced by a universal population model. This structure enables hierarchical models to effectively capture individual variations among drivers while reducing overfitting risks. It is important to clarify that the variance of $\boldsymbol{\theta}_d$ in the hierarchical model is greater than in the pooled model and smaller than in the unpooled model. The hierarchical model's behavior reflects the degree of this variance: with greater variance, it mirrors an unpooled model, and with minimal variance, it resembles a pooled model. Understanding this variance is crucial for fine-tuning hierarchical model hyperparameters.

\begin{figure*}[t]
    \centering
    \subfigure[The pooled B-IDM.]{\centering
    \resizebox{!}{3.0cm}{\tikzset{every picture/.style={line width=0.75pt}} 

\begin{tikzpicture}[x=0.75pt,y=0.75pt,yscale=-1,xscale=1]

\node [obs, draw=black,inner sep=0pt,minimum size=0.75cm] (o) at (-25,20) {${a}_d^{(t)}$};

\node [obs, draw=blue,inner sep=0pt,minimum size=0.75cm] (x) at (-25,-40) {$\boldsymbol{i}_d^{(t)}$};

\node [circle,draw=blue,fill=white, inner sep=0pt,minimum size=0.75cm] (theta) at (-110,20) {$\boldsymbol{\theta}$};

\node [circle,draw=black] (tau) at (60,20) {$\sigma_{\epsilon}$};


\path [draw,->, color=blue] (x) edge (o);
\path [draw,->, color=blue] (theta) edge (o);
\path [draw,->] (tau) edge (o);

\node [text width=3.cm] (t) at (-10,43) {\scriptsize{Time $t \in \{1,\dots, T_d\}$}};
\node [text width=3.cm] (driver) at (-23,62) {\scriptsize{Drivers $d \in \{1,\dots, D\}$}};
\node [text width=1cm] (placeholder) at (5,50) {};
\node [text width=1cm] (placeholder3) at (-55,-55) {};
\node [text width=1cm] (placeholder5) at (0,-40) {};
\node [text width=1cm, color=blue] (f) at (-40,-16) {${\text{IDM}}$};

\plate [color=black] {t_level} {(x)(o)(f)(placeholder5)} { };
\plate [color=black] {d_level} {(x)(o)(f)(placeholder3)(placeholder)} { };

\end{tikzpicture}}}
    \subfigure[The hierarchical B-IDM.]{\centering
    \resizebox{!}{3.0cm}{\tikzset{every picture/.style={line width=0.75pt}} 

\begin{tikzpicture}[x=0.75pt,y=0.75pt,yscale=-1,xscale=1]
\node [obs, draw=black,inner sep=0pt] (o) at (-30,20) {${a}_{d}^{(t)}$};
\node [obs, draw=blue,inner sep=0pt,minimum size=0.75cm] (x) at (-30,-40) {$\boldsymbol{i}_d^{(t)}$};
\node [circle,draw=blue,fill=white, inner sep=0pt,minimum size=0.75cm] (theta) at (-100,20) {$\boldsymbol{\theta}_d$};

\node [circle,draw=black] (Sigma) at (-150,0) {$\boldsymbol{\Sigma}$};
\node [circle,draw=black] (mu) at (-150,40) {$\boldsymbol{\theta}$};
\node [circle,draw=black] (tau) at (45,20) {$\sigma_{\epsilon,d}$};
\node [circle,draw=black] (tau2) at (95,20) {$\sigma_{\epsilon}$};
\node [circle,draw=black] (sigma_0) at (-150,-45) {$\boldsymbol{\sigma}$};
\path [draw,->, color=black] (sigma_0) edge (Sigma);

\path [draw,->] (tau2) edge (tau);
\path [draw,->] (Sigma) edge (theta);
\path [draw,->, color=blue] (x) edge (o);
\path [draw,->, color=blue] (theta) edge (o);
\path [draw,->] (tau) edge (o);

\path [draw,->] (mu) edge (theta);
\node [text width=3.cm] (driver) at (-62,62) {\scriptsize{Drivers $d \in \{1,\dots, D\}$}};
\node [text width=3.cm] (t) at (-15,42) {\scriptsize{Time $t \in \{1,\dots, T_d\}$}};
\node [text width=1cm] (placeholder) at (45,50) {};
\node [text width=1cm] (placeholder3) at (-95,-55) {};
\node [text width=1cm] (placeholder5) at (-6,-40) {};
\node [text width=1cm, color=blue] (f) at (-45,-16) {${\text{IDM}}$};

\plate [color=black] {t_level} {(x)(o)(f)(placeholder5)} { };
\plate [color=black] {driver_level} {(x)(o)(theta)(placeholder)(placeholder3)} { };

\end{tikzpicture}}
    }
    \subfigure[The unpooled B-IDM.]{\centering
    \resizebox{!}{3.0cm}{\tikzset{every picture/.style={line width=0.75pt}} 

\begin{tikzpicture}[x=0.75pt,y=0.75pt,yscale=-1,xscale=1]

\node [obs, draw=black,inner sep=0pt] (o) at (-25,20) {${a}_d^{(t)}$};

\node [obs, draw=blue,inner sep=0pt,minimum size=0.75cm] (x) at (-25,-40) {$\boldsymbol{i}_d^{(t)}$};

\node [circle,draw=blue,fill=white, inner sep=0pt,minimum size=0.75cm] (theta) at (-110,20) {$\boldsymbol{\theta}_d$};

\node [circle,draw=black] (tau) at (45,20) {$\sigma_{\epsilon,d}$};


\path [draw,->, color=blue] (x) edge (o);
\path [draw,->, color=blue] (theta) edge (o);
\path [draw,->] (tau) edge (o);

\node [text width=3.cm] (driver) at (-23,62) {};
\node [text width=3.cm] (t) at (-14,42) {\scriptsize{Time $t \in \{1,\dots, T_d\}$}};
\node [text width=3.cm] (driver) at (-70,62) {\scriptsize{Drivers $d \in \{1,\dots, D\}$}};
\node [text width=1cm] (placeholder) at (40,50) {};
\node [text width=1cm] (placeholder3) at (-105,-55) {};
\node [text width=1cm] (placeholder5) at (-6,30) {};
\node [text width=1cm, color=blue] (f) at (-45,-16) {${\text{IDM}}$};

\plate [color=black] {t_level} {(x)(o)(f)(placeholder5)} { };

\plate [color=black] {driver_level} {(x)(o)(theta)(placeholder)(placeholder3)} { };

\end{tikzpicture}}
    }\\
    \subfigure[The pooled MA-IDM.]{\centering
    \resizebox{!}{3.0cm}{\tikzset{every picture/.style={line width=0.75pt}} 

\begin{tikzpicture}[x=0.75pt,y=0.75pt,yscale=-1,xscale=1]

\node [obs, draw=black,inner sep=0pt,minimum size=0.75cm] (o) at (-25,20) {$\boldsymbol{a}_d$};

\node [obs, draw=blue,inner sep=0pt,minimum size=0.75cm] (x) at (-25,-40) {$\boldsymbol{i}_d$};

\node [circle,draw=blue,fill=white, inner sep=0pt,minimum size=0.75cm] (theta) at (-110,20) {$\boldsymbol{\theta}$};


\node [circle,draw=red,fill=white, inner sep=0pt,minimum size=0.75cm] (ell) at (60,40) {$\ell$};
\node [circle,draw=red,fill=white, inner sep=0pt,minimum size=0.75cm] (sigma) at (60,0) {$\sigma_k$};



\path [draw,->, color=blue] (x) edge (o);
\path [draw,->, color=blue] (theta) edge (o);
\path [draw,->, color=red] (ell) edge (o);
\path [draw,->, color=red] (sigma) edge (o);

\node [text width=3.cm] (driver) at (-23,62) {\scriptsize{Drivers $d \in \{1,\dots, D\}$}};
\node [text width=1cm] (placeholder) at (5,50) {};
\node [text width=1cm] (placeholder3) at (-55,-55) {};
\node [text width=1cm] (placeholder5) at (28,-40) {};
\node [text width=1cm, color=blue] (f) at (-45,-16) {${\text{IDM}}$};
\node [text width=1cm, color=red] (GP) at (30,20) {GP};

\plate [color=black] {d_level} {(x)(o)(f)(placeholder3)(placeholder)} { };

\end{tikzpicture}}}
    \subfigure[The hierarchical MA-IDM.]{\centering
    \resizebox{!}{3.0cm}{\tikzset{every picture/.style={line width=0.75pt}} 

\begin{tikzpicture}[x=0.75pt,y=0.75pt,yscale=-1,xscale=1]
\node [obs, draw=black,inner sep=0pt,minimum size=.75cm] (o) at (-30,20) {$\boldsymbol{a}_{d}$};
\node [obs, draw=blue,inner sep=0pt,minimum size=0.75cm] (x) at (-30,-40) {$\boldsymbol{i}_{d}$};
\node [circle,draw=blue,fill=white, inner sep=0pt,minimum size=0.75cm] (theta) at (-100,20) {$\boldsymbol{\theta}_d$};

\node [circle,draw=black] (mu) at (-150,40) {$\boldsymbol{\theta}$};
\node [circle,draw=black] (Sigma) at (-150,0) {$\boldsymbol{\Sigma}$};
\node [circle,draw=black] (sigma_0) at (-150,-45) {$\boldsymbol{\sigma}$};
\path [draw,->, color=black] (sigma_0) edge (Sigma);

\node [circle,draw=red,fill=white, inner sep=0pt,minimum size=0.75cm] (ell) at (40,40) {$\ell_d$};
\node [circle,draw=red,fill=white, inner sep=0pt,minimum size=0.75cm] (sigma) at (40,0) {$\sigma_{k,d}$};

\node [circle,draw=red,fill=white, inner sep=0pt,minimum size=0.75cm] (ell0) at (90,40) {$\ell$};
\node [circle,draw=red,fill=white, inner sep=0pt,minimum size=0.75cm] (sigma0) at (90,0) {$\sigma_{k}$};

\path [draw,->] (Sigma) edge (theta);
\path [draw,->, color=blue] (x) edge (o);
\path [draw,->, color=blue] (theta) edge (o);
\path [draw,->, color=red] (ell) edge (o);
\path [draw,->, color=red] (ell0) edge (ell);
\path [draw,->, color=red] (sigma) edge (o);
\path [draw,->, color=red] (sigma0) edge (sigma);
\path [draw,->] (mu) edge (theta);

\node [text width=3.cm] (driver) at (-60,62) {\scriptsize{Drivers $d \in \{1,\dots, D\}$}};
\node [text width=1cm] (placeholder) at (35,50) {};
\node [text width=1cm] (placeholder3) at (-95,-55) {};
\node [text width=1cm] (placeholder5) at (0,-40) {};
\node [text width=1cm, color=red] (GP) at (25,20) {GP};
\node [text width=1cm, color=blue] (f) at (-45,-16) {${\text{IDM}}$};

\plate [color=black] {driver_level} {(x)(o)(theta)(placeholder)(placeholder3)} { };

\end{tikzpicture}}
    }
    \subfigure[The unpooled MA-IDM.]{\centering
    \resizebox{!}{3.0cm}{\tikzset{every picture/.style={line width=0.75pt}} 

\begin{tikzpicture}[x=0.75pt,y=0.75pt,yscale=-1,xscale=1]

\node [obs, draw=black,inner sep=0pt,minimum size=0.75cm] (o) at (-30,20) {$\boldsymbol{a}_d$};

\node [obs, draw=blue,inner sep=0pt,minimum size=0.75cm] (x) at (-30,-40) {$\boldsymbol{i}_d$};

\node [circle,draw=blue,fill=white, inner sep=0pt,minimum size=0.75cm] (theta) at (-110,20) {$\boldsymbol{\theta}_d$};


\node [circle,draw=red,fill=white, inner sep=0pt,minimum size=0.75cm] (ell) at (40,40) {$\ell_d$};
\node [circle,draw=red,fill=white, inner sep=0pt,minimum size=0.75cm] (sigma) at (40,0) {$\sigma_{k,d}$};


\node [text width=3.cm] (driver) at (-70,62) {};


\path [draw,->, color=blue] (x) edge (o);
\path [draw,->, color=blue] (theta) edge (o);
\path [draw,->, color=red] (ell) edge (o);
\path [draw,->, color=red] (sigma) edge (o);

\node [text width=3.cm] (driver) at (-66,62) {\scriptsize{Drivers $d \in \{1,\dots, D\}$}};
\node [text width=1cm] (placeholder) at (35,50) {};
\node [text width=1cm] (placeholder3) at (-100,-55) {};
\node [text width=1cm] (placeholder5) at (28,-40) {};
\node [text width=1cm, color=blue] (f) at (-45,-16) {${\text{IDM}}$};
\node [text width=1cm, color=red] (GP) at (25,20) {GP};
\plate [color=black] {d_level} {(x)(o)(f)(placeholder3)(placeholder)} { };

\end{tikzpicture}}}
    \caption{Probabilistic graphical models.} \label{4PGM}
\end{figure*}
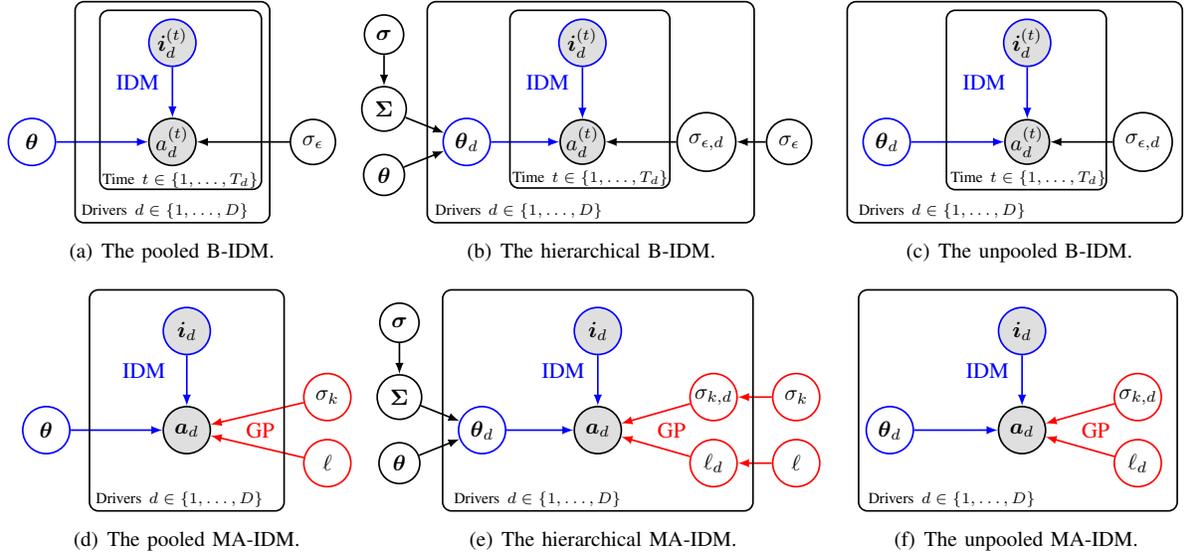

\section{Bayesian Calibration of Car Following Model}\label{method}
In this section, we develop a novel Bayesian calibration approach and apply it to identify the IDM parameters. Specifically, we present the formulations of four models, i.e., the pooled and hierarchical models of the B-IDM and the MA-IDM, respectively. The corresponding probabilistic graphical models are shown in Fig.~\ref{4PGM}.

\subsection{Pooled and Hierarchical B-IDM}
Firstly, we introduce the form of the pooled B-IDM: $\forall (t,d)\in\{(t,d)\}_{t=t_0, d=1}^{t_0+(T-1)\Delta t,D}$, we have
\begin{subequations}\label{Eq5}
\begin{align}
    &\ln(\boldsymbol{\theta}) \sim \mathcal{N}(\ln(\boldsymbol{\theta}_{\text{rec}}), \boldsymbol{\Sigma}_0) \in \mathbb{R}^{5},\\
    &\ln(\sigma_{\epsilon}) \sim \mathcal{N}(\ln(\mu_{\sigma}),\sigma_{0}^2) \in \mathbb{R}, \\
    & \mathbf{for\,driver}\,d = 1,\ldots,D:\nonumber\\
    & \quad \quad \mathbf{for\,time}\, t=t_0,\ldots, t_0+(T_d-1)\Delta t:\nonumber\\
    &\quad \quad \quad \quad {a}_d^{(t)}|\boldsymbol{h}_d^{(t)},\boldsymbol{\theta}\, \stackon{$\sim$}{\tiny{\textit{i.i.d.}}}\, \mathcal{N}(a_{\text{IDM}}, \sigma_{\epsilon}^2) \in \mathbb{R},\label{v_pool_B-IDM}
\end{align}
\end{subequations}
where $\boldsymbol{\theta}_{\text{rec}}$, $\boldsymbol{\Sigma}_0$, $\mu_{\sigma}$ and $\sigma_{0}$ are all the manually set hyperparameters. In particular, the IDM priors are set by mainly following the recommended values $\boldsymbol{\theta}_{\text{rec}}=[33.3, 2.0, 1.6, 1.5, 1.67]$ in \cite{treiber2000congested, punzo2014we}. Note that in the literature \cite{abodo2019strengthening}, some samples drawn from Markov chains are unrealistic with negative values. To avoid this, {e.g., as the model described in Eq.~\eqref{Eq5}, we assume log-normal distributions to project the variables to the set of non-negative real numbers $\mathbb{R}^+$}. With such settings, one can ensure that the corresponding parameters are positive.

The capability of the pooled model is limited since it can only capture the general driving styles of a ``population driver.'' To characterize and model the heterogeneity in human driving behavior, one can learn different parameters for different drivers, which translates to introducing heterogeneity at the level of the driver population \cite{treiber2013traffic}. Here we develop a hierarchical B-IDM that explicitly models the general driving behaviors at the population level, then depicts heterogeneous behaviors at the individual level, described as
\begin{subequations}
\begin{align}
    &\boldsymbol{\sigma} \sim \mathrm{Exp}(\lambda) \in \mathbb{R}^{5},\label{6a}\\
    &\boldsymbol{\Sigma}  \sim \mathrm{LKJCholeskyCov}(\eta, \boldsymbol{\sigma}) \in \mathbb{R}^{5\times 5}, \\
    &\ln(\boldsymbol{\theta})  \sim \mathcal{N}(\ln(\boldsymbol{\theta}_{\text{rec}}), \boldsymbol{\Sigma}_0) \in \mathbb{R}^{5},\\
    &\sigma_{\epsilon}  \sim \mathrm{Exp}(\lambda_{\epsilon}) \in \mathbb{R}, \\
    & \mathbf{for\,driver}\,d = 1,\ldots,D:\nonumber\\
    &\quad \quad \ln(\boldsymbol{\theta}_d)  \sim \mathcal{N}(\ln(\boldsymbol{\theta}), \boldsymbol{\Sigma}) \in \mathbb{R}^{5}, \label{individual_B}\\
    &\quad \quad \ln(\sigma_{\epsilon,d})  \sim \mathcal{N}(\ln(\sigma_{\epsilon}),\sigma_{\sigma}^2) \in \mathbb{R}, \\
    & \quad \quad \mathbf{for\,time}\, t=t_0,\ldots, t_0+(T_d-1)\Delta t:\nonumber\\
    &\quad \quad \quad \quad {a}_d^{(t)}|\boldsymbol{h}_d^{(t)},\boldsymbol{\theta}_d \,\stackon{$\sim$}{\tiny{\textit{i.i.d.}}}\, \mathcal{N}(a_{\text{IDM}}, \sigma_{\epsilon,d}^2) \in \mathbb{R},
\end{align}
\end{subequations}
where $\mathrm{Exp}(\lambda)$ represents the exponential distribution with rate $\lambda$, and LKJ stands for the LKJ distribution developed by Lewandowski, Kurowicka, and Joe \cite{lewandowski2009generating}. The term ``LKJCholeskyCov'' represents a distribution over Cholesky decomposed covariance matrices, such that the underlying correlation matrices follow an LKJ distribution and the standard deviations follow an exponential distribution, as specified by PyMC \cite{salvatier2016probabilistic}. In practice, the LKJCholeskyCov distribution is a more robust prior for covariance matrix than the inverse-Wishart distribution \cite{alvarez2014bayesian}. Here we put the same prior in Eq.~\eqref{6a} for the five IDM parameters. However, note that the scales of these parameters are different. Thus, it is necessary to perform standardization. 

\subsection{Pooled and Hierarchical MA-IDM}\label{unbiased_model}
For daily driving tasks, temporally consistent inertia of actions/motions is one of the vital driving profiles for human drivers. This phenomenon is termed as \textit{driving persistence} in \cite{treiber2013traffic}. For instance, if drivers slam on their brakes too hard at a certain timestamp, they will likely persist in a vast deceleration in the next seconds. However, in most stochastic models, as well as B-IDM introduced above, the persistence of the acceleration noise is ignored, and its time dependence is incorrectly modeled by white noise (see Eq.~\eqref{iid_a}). To make this point clear, we reformulate Eq.~\eqref{iid_a} as
\begin{equation}\label{reformulated_iid_a}
    a_d^{(t)} = a_{\mathrm{IDM},d}^{(t)} + \epsilon_d^{(t)},\ \epsilon_d^{(t)}\,\stackon{$\sim$}{\tiny{\textit{i.i.d.}}}\,\mathcal{N}(0,\sigma_{\epsilon,d}^2).
\end{equation}

However, the developed models and the identified parameters are only valid if the errors conform to the \textit{i.i.d.} assumption since the temporal correlation in the residuals leads to biased parameter estimation. In the following, we will derive an unbiased Bayesian model in the pooled and hierarchical form by modeling the time-dependent stochastic error term with Gaussian processes.

Given the nature of time-series autocorrelation, we construct the MA-IDM by modeling $\epsilon_d^{(t)}$ in Eq.~\eqref{reformulated_iid_a} with Gaussian processes such that
\begin{align}\label{MA_idm_eq}
    a_d^{(t)} = a_{\mathrm{IDM},d}^{(t)} + a_{\mathrm{GP},d}^{(t)},
\end{align}
with $a_{\mathrm{GP},d}^{(t)}=\epsilon_d^{(t)}$ depicts a serial correlated error term, written as $a_{\mathrm{GP},d}^{(t)}\sim \mathcal{GP}(0, k_d(t,t'))$, where $k$ is the kernel function. The prior distribution of $\boldsymbol{a}_{\mathrm{GP},d}=\left[a_{\mathrm{GP},d}^{(t_0)},\ldots,a_{\mathrm{GP},d}^{(t_0+(T_d-1)\Delta t)}\right]^{\top}$ is usually set as a zero-mean Gaussian distribution $\mathcal{N}(\boldsymbol{0},\boldsymbol{K}_d)$, where $\boldsymbol{K}_d=[k_d(i,j)]$ is a ${T_d\times T_d}$ positive definite covariance matrix. To characterize the positive temporal autocorrelation in the error process, we introduce a squared-exponential (SE) kernel function.
\begin{equation}
    k_d(t,t') = \sigma_{k,d}^2\,\mathrm{exp}\left( -\frac{|t-t'|^2}{2 \ell_d^2} \right),
\end{equation}
where $\ell_d$ is the lengthscale and $\sigma_{k,d}^2$ represents the noise variance for driver $d$, respectively. The lengthscale $\ell_d$ determines how the correlation decays with $|t-t'|$ as the distance between the two consecutive time steps---a smaller lengthscale corresponds to a faster decaying correlation. Note that one can also choose other kernel functions such as the Mat\'{e}rn 5/2 kernel. Combining $\boldsymbol{a}_{\mathrm{IDM},d}$ and $\boldsymbol{a}_{\mathrm{GP},d}$, we have
\begin{align}\label{a_with_gp}
    \boldsymbol{a}_d|\boldsymbol{h}_d,\boldsymbol{\theta}_d \sim \mathcal{N}(\boldsymbol{a}_{\mathrm{IDM},d}, \boldsymbol{K}_d) \in \mathbb{R}^{T_d}.
\end{align}

Therefore, for the hierarchical MA-IDM model, we have
\begin{subequations}
\begin{align}
    &\boldsymbol{\sigma} \sim \mathrm{Exp}(\lambda) \in \mathbb{R}^{5},\label{15a}\\
    &\boldsymbol{\Sigma}  \sim \mathrm{LKJCholeskyCov}(\eta, \boldsymbol{\sigma}) \in \mathbb{R}^{5\times 5}, \\
    &\ln(\boldsymbol{\theta})  \sim \mathcal{N}(\ln(\boldsymbol{\theta}_{\text{rec}}), \boldsymbol{\Sigma}_0) \in \mathbb{R}^{5},\\
    &\sigma_{k} \sim \mathrm{Exp}(\lambda_{k}) \in \mathbb{R}, \\
    &\ln(\ell)  \sim \mathcal{N}(\ln(\mu_{\ell}),\sigma_{0}^2) \in \mathbb{R}, \\
    & \mathbf{for\,driver}\,d = 1,\ldots,D:\nonumber\\
    & \quad \quad \ln(\boldsymbol{\theta}_d)  \sim \mathcal{N}(\ln(\boldsymbol{\theta}), \boldsymbol{\Sigma}) \in \mathbb{R}^{5}, \label{individual_GP}\\
    &\quad \quad \ln(\sigma_{k,d})  \sim \mathcal{N}(\ln(\sigma_{k}),\sigma_{\sigma}^2) \in \mathbb{R}, \\
    &\quad \quad \ln(\ell_{d})  \sim \mathcal{N}(\ln(\ell),\sigma_{\ell}^2) \in \mathbb{R}, \\
    &\quad \quad \boldsymbol{a}_d|\boldsymbol{h}_d,\boldsymbol{\theta}_d\, \stackon{$\sim$}{\tiny{\textit{i.i.d.}}}\, \mathcal{N}\left(\boldsymbol{a}_{\text{IDM}}, \boldsymbol{K}_d\right) \in \mathbb{R}^{T_d}.\label{v_pool_MA-IDM}
\end{align}
\end{subequations}

Note that, different from Eq.~\eqref{v_pool_B-IDM}, here we use a multivariate normal distribution to model the vector-form actions in Eq.~\eqref{v_pool_MA-IDM}. The pooled model can be simply realized by assuming the same $\boldsymbol{\theta}$, $\sigma_k$, and $\ell$ for all drivers:
\begin{subequations}
\begin{align}
    &\ln(\boldsymbol{\theta}) \sim \mathcal{N}(\ln(\boldsymbol{\theta}_{\text{rec}}), \boldsymbol{\Sigma}_0) \in \mathbb{R}^{5},\\
    &\ln(\sigma_{k}) \sim \mathcal{N}(\ln(\mu_{\sigma}),\sigma_{0}^2) \in \mathbb{R}, \\
    &\ln(\ell)  \sim \mathcal{N}(\ln(\mu_{\ell}),\sigma_{0}^2) \in \mathbb{R}, \\
    & \mathbf{for\,driver}\,d = 1,\ldots,D:\nonumber\\
    &\quad \quad \boldsymbol{a}_d|\boldsymbol{h}_d,\boldsymbol{\theta}\, \stackon{$\sim$}{\tiny{\textit{i.i.d.}}}\, \mathcal{N}\left(\boldsymbol{a}_{\text{IDM}}, \boldsymbol{K}\right) \in \mathbb{R}^{T_d}.
\end{align}
\end{subequations}

\subsection{Calibrate MA-IDM and B-IDM}
We first introduce the calibration for MA-IDM. According to Eq.~\eqref{simulation_update}, we update the speed and location as
\begin{align}
    \left[\begin{array}{c}
          v_d^{(t+\Delta t)} \\ x_d^{(t+\Delta t)}
    \end{array}\right] =&  \left[\begin{array}{c}
        m_v^{(t)}  \\
        m_x^{(t)}
    \end{array}\right] + \underbrace{\left[\begin{array}{c}
         \Delta t  \\
         \frac{1}{2}\Delta t^2
    \end{array}\right]}_{\coloneqq\boldsymbol{C}} a_{\mathrm{GP}_d}^{(t)},\\
    \left[\begin{array}{c}
        m_v^{(t)}  \\
        m_x^{(t)}
    \end{array}\right] =& \left[\begin{array}{c}
          v_d^{(t)} + a_{\mathrm{IDM},d}^{(t)}\Delta t \\ x_d^{(t)} +v_d^{(t)}\Delta t + \frac{1}{2}a_{\mathrm{IDM},d}^{(t)}\Delta t^2
    \end{array}\right].
\end{align}
which is equivalent to the vector form expression
\begin{align}
    \left[\begin{array}{c}
          \boldsymbol{v}_d \\ \boldsymbol{x}_d
    \end{array}\right]\,\sim\, \mathcal{N}\left(\left[\begin{array}{c}
        \boldsymbol{m}_v  \\ \boldsymbol{m}_x
    \end{array}\right], \underbrace{\left[\begin{array}{cc}
        \Delta t^2 & \frac{1}{2}\Delta t^3 \\
        \frac{1}{2}\Delta t^3 &  \frac{1}{4}\Delta t^4
    \end{array}\right]}_{=\boldsymbol{C}\boldsymbol{C}^\top} \otimes \boldsymbol{K}_d\right),
\end{align}
where $\otimes$ represents the Kronecker product, the speeds are written in the vector form as $\boldsymbol{v}_d  = [v_d^{(t+\Delta t)}, \dots, v_d^{(t+T\Delta t)}]^\top\in \mathbb{R}^{T}$, and likewise for $\boldsymbol{x}_d$, $\boldsymbol{m}_v$, and $\boldsymbol{m}_x$. It should be noted that data duration $T$ for each driver (i.e., trajectory) should be different; for simplicity, we remove the driver index $d$ in $T$.

Then, as the data we observed are usually polluted by noise, we can derive a likelihood that is jointly measured on both the speed and spacing data by considering observation noise
\begin{align}
    \left[\begin{array}{c}
          \hat{\boldsymbol{v}}_d \\ \hat{\boldsymbol{x}}_d
    \end{array}\right]\,\sim\, \mathcal{N}\left(\left[\begin{array}{c}
        \boldsymbol{m}_v  \\ \boldsymbol{m}_x
    \end{array}\right],  \boldsymbol{C}\boldsymbol{C}^\top \otimes \boldsymbol{K}_d+\left[\begin{array}{cc}
         \sigma_v^2 & 0 \\
         0 & \sigma_x^2
    \end{array}\right] \otimes \boldsymbol{I}\right), \label{joint_llh}
\end{align}
where $\hat{\boldsymbol{v}}_d$ and $\hat{\boldsymbol{x}}_d$ represent the observed data, $\boldsymbol{I}\in \mathbb{R}^{T\times T}$ is the identity matrix, $\sigma_v$ and $\sigma_x$ are the observation noise level of the speed and location data, respectively. A large noise variance indicates that the information carried by the data is unreliable, and thus has limited contributions to the likelihood. To let our model automatically identify different noise levels, it is necessary to add two more priors on $\sigma_v$ and $\sigma_x$
\begin{align}
    \sigma_{v} & \sim \mathrm{Exp}(\lambda_{v}) \in \mathbb{R}^+,\\
    \sigma_{x} & \sim \mathrm{Exp}(\lambda_{x}) \in \mathbb{R}^+.
\end{align}
Note that when purely calibrating on either the speed data or the spacing data, one can easily derive the following forms as the special cases of Eq.~\eqref{joint_llh}:
\begin{align}
    \hat{\boldsymbol{v}}_d \sim&\,\mathcal{N}(\boldsymbol{m}_v, \Delta t^2\boldsymbol{K}_d+\sigma_v^2\boldsymbol{I}),\\
    \hat{\boldsymbol{x}}_d \sim&\,\mathcal{N}(\boldsymbol{m}_x, \frac{1}{4}\Delta t^4\boldsymbol{K}_d+\sigma_x^2\boldsymbol{I}).
\end{align}
For the calibration of the B-IDM, we can simply replace $\boldsymbol{K}_d$ with $\sigma_{\epsilon,d}^2\boldsymbol{I}$ in the above derivation. It is worth noting that in this case the likelihood can be factorized for each time stamp, i.e.,
\begin{align}
    \hat{{v}}_d^{(t)} \sim&\,\mathcal{N}({m}_v^{(t)}, \Delta t^2\sigma_{\epsilon,d}^2+\sigma_v^2), \label{eq_bidm_llh_v}\\
    \hat{{x}}_d^{(t)} \sim&\,\mathcal{N}({m}_x^{(t)}, \frac{1}{4}\Delta t^4\sigma_{\epsilon,d}^2+\sigma_x^2).\label{eq_bidm_llh_x}
\end{align}

\begin{figure*}[!t]
    \centering
    \subfigure[Trajectory visualization.]{
        \centering
        \includegraphics[width=0.48\textwidth]{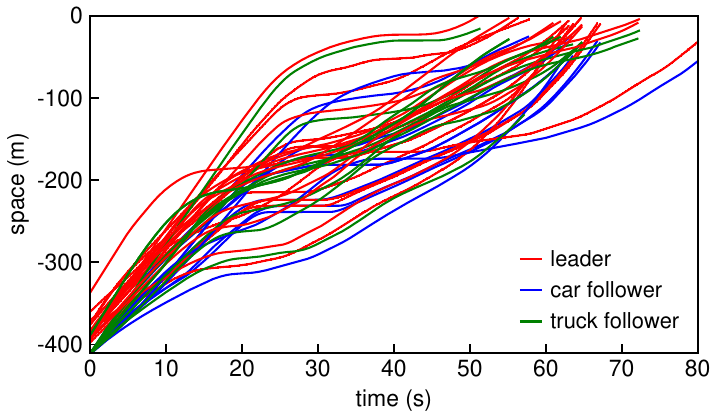}
    }
    \subfigure[Empirical histogram.]{
        \centering
        \includegraphics[width=0.48\textwidth]{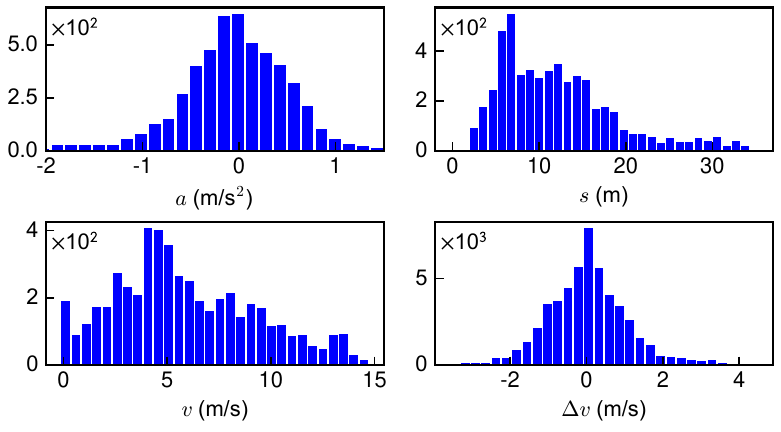}
    }
    \caption{Visualizations of the {$20$} selected leader-follower pairs. }
    \label{fig:Visualizations}
\end{figure*}

\section{Experiments}\label{experiments}
In this section, we assess the performance of our proposed car-following models by calibrating on naturalistic human-driving car-following data.

\subsection{Experimental Settings}
\subsubsection{Dataset}
The behavior of car-following models can be affected and contaminated by noise in empirical data \cite{montanino2015trajectory}. Therefore, selecting an appropriate dataset can keep the model performance away from unsatisfied data quality and relieving us from the redundant work of data cleaning. In this paper, we adopt the HighD dataset \cite{highDdataset}---a high-resolution trajectory data collected in Germany using drones. Compared to the widely used NGSIM data set \cite{punzo2011assessment}, HighD benefits from advanced computer vision techniques and reliable data-capture methods.

The HighD dataset has $60$ video recordings, logged with the sampling frequency of $25$ Hz on several German highway sections with a length of $420$ m. The original dataset is downsampled to a smaller set with a sampling frequency of 5 Hz, achieved by uniformly selecting every 5th sample. In each recording, the trajectories, speeds, and accelerations of two types of vehicles (car and truck) are measured and filtered. We follow the same data processing procedures as in \cite{zhang2021spatiotemporal} to transform the data into a new coordinate system.

\subsubsection{Car-Following Data Extraction}
According to \cite{punzo2014we}, long trajectories are preferred for robust estimation of IDM parameters. Thus, we first discard data with a car-following duration less than a certain threshold $t_0=50$ s to keep informative leader-follower pairs \cite{zhang2023interactive}. Then we randomly select $20$ leader-follower pairs from these data, and $10$ pairs for each type of vehicle, respectively. Fig.~\ref{fig:Visualizations} shows examples of the trajectories used for calibration. It should be noted that here we only consider the heterogeneity of the followers; if interested, readers can also investigate different combinations of more types of vehicles for both leaders and followers, e.g., as discussed in \cite{liu2016modeling}.
The computational complexity of learning the GP for driver $d$ is $\mathcal{O}(T_d^3)$. This becomes particularly problematic for MCMC when $T_d$ is large. To reduce the computational cost in estimating the model, we divide the whole sequence into several segments with a fixed window size and consider each segment as an independent observation. In the experiment, we set the segment length to $4$ s so that the covariance matrix $\boldsymbol{K}_d$ has a fixed size of $20\times 20$.

\subsubsection{MCMC Settings}
In this work, we use PyMC \cite{salvatier2016probabilistic} to sample from the posteriors. Specifically, we adopt one kind of Hamiltonian Monte Carlo (HMC) method, i.e., the No-U-Turn Sampler (NUTS) \cite{hoffman2014no}, and we set the burn-in steps as $5000$ to ensure the detailed balance condition holds for the Markov chains. We next describe the settings for the hyperparameters of priors. We set $\lambda=100$ and $\eta=2$ for the LKJCholeskyCov distribution. Looking into Eqs.~\eqref{reformulated_iid_a} and \eqref{MA_idm_eq}, one can see a potential identifiability issue as the acceleration can be explained by both the IDM function and the error term. To avoid this issue, we set strong priors for the scale of the error variance, i.e., $\lambda_\epsilon=\lambda_k=5000$ for the exponential distributions; $\lambda_x=1000$ and $\lambda_v=1\times10^5$ for the observation noises. At the individual level, the priors of $\sigma_{k,d}$ and $\sigma_{\epsilon,d}$ are set with $\sigma_\sigma=0.05$. The hyperparameters for the log-normal distributions of $\ell$ are flexible, we set $\sigma_0=0.2$ and $\sigma_\ell=0.05$, and one can tune the prior to ensure that it is a positive value falling in a reasonable range. It should be noted that $\lambda$ controls the variance of the population-level IDM parameters. Guided by the illustration in Fig.~\ref{pooling_techinique}, for the exponential distributions, setting a large $\lambda$ results in a small variance, thus the hierarchical model tends to perform like a pooled model and vice versa.

\subsubsection{Reproducibility}
All codes for reproducing the experiment results reported in this paper are available at \url{https://github.com/Chengyuan-Zhang/IDM_Bayesian_Calibration}.

\begin{figure*}[t]
    \centering
    \subfigure[Hierarchical B-IDM posteriors of car $\# 273$.]{
        \centering
        \includegraphics[width=0.483\textwidth]{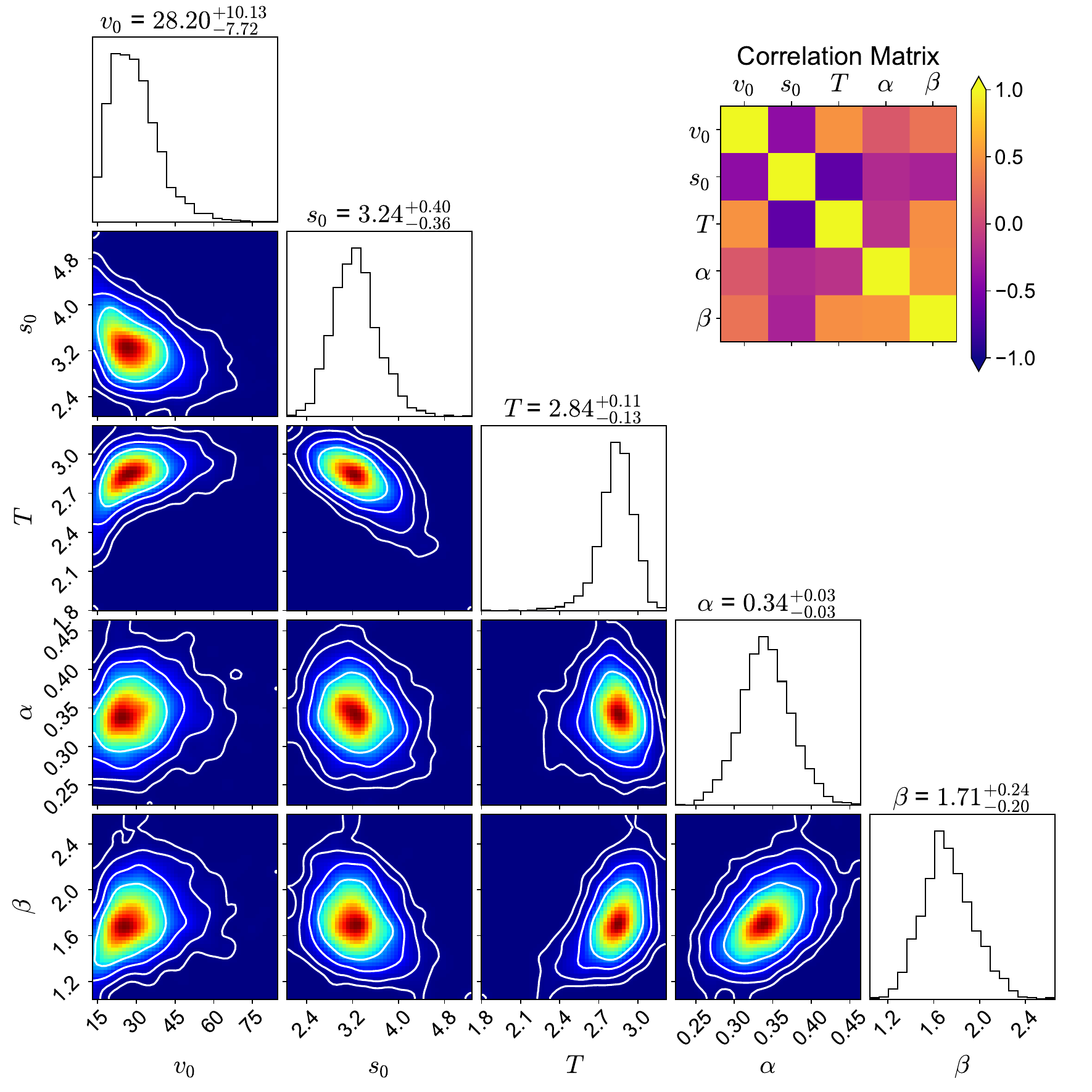}
    }
    \subfigure[Hierarchical MA-IDM posteriors of car $\# 273$.]{
        \centering
        \includegraphics[width=0.483\textwidth]{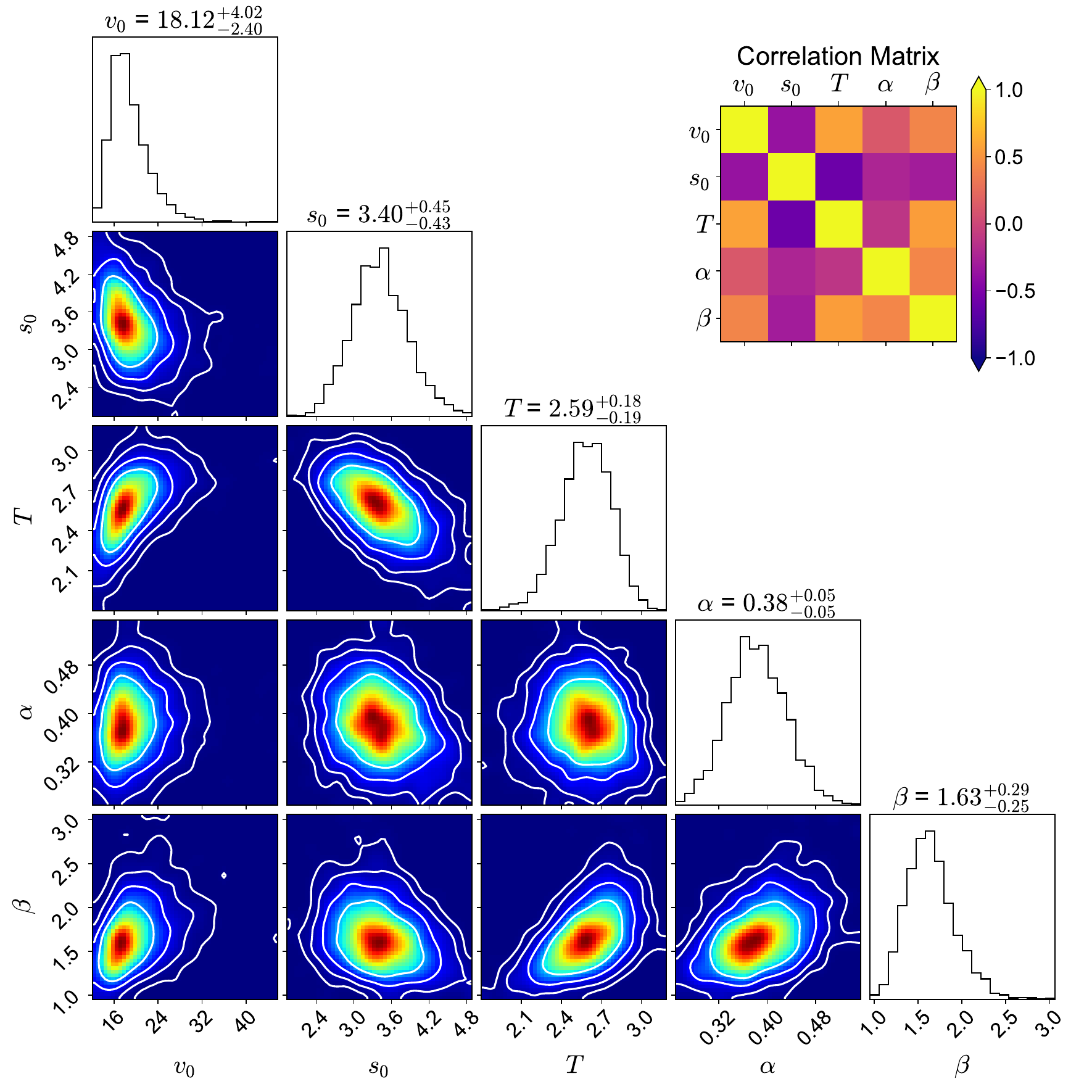}
    }
    \caption{The posterior distributions and the correlation matrices of the parameters of a car driver $\#273$ in the hierarchical B-IDM and MA-IDM.}
    \label{fig:IDM_two_driver355}
\end{figure*}

\begin{figure*}[t]
    \centering
    \subfigure[Hierarchical B-IDM posteriors of truck $\# 211$.]{
        \centering
        \includegraphics[width=0.483\textwidth]{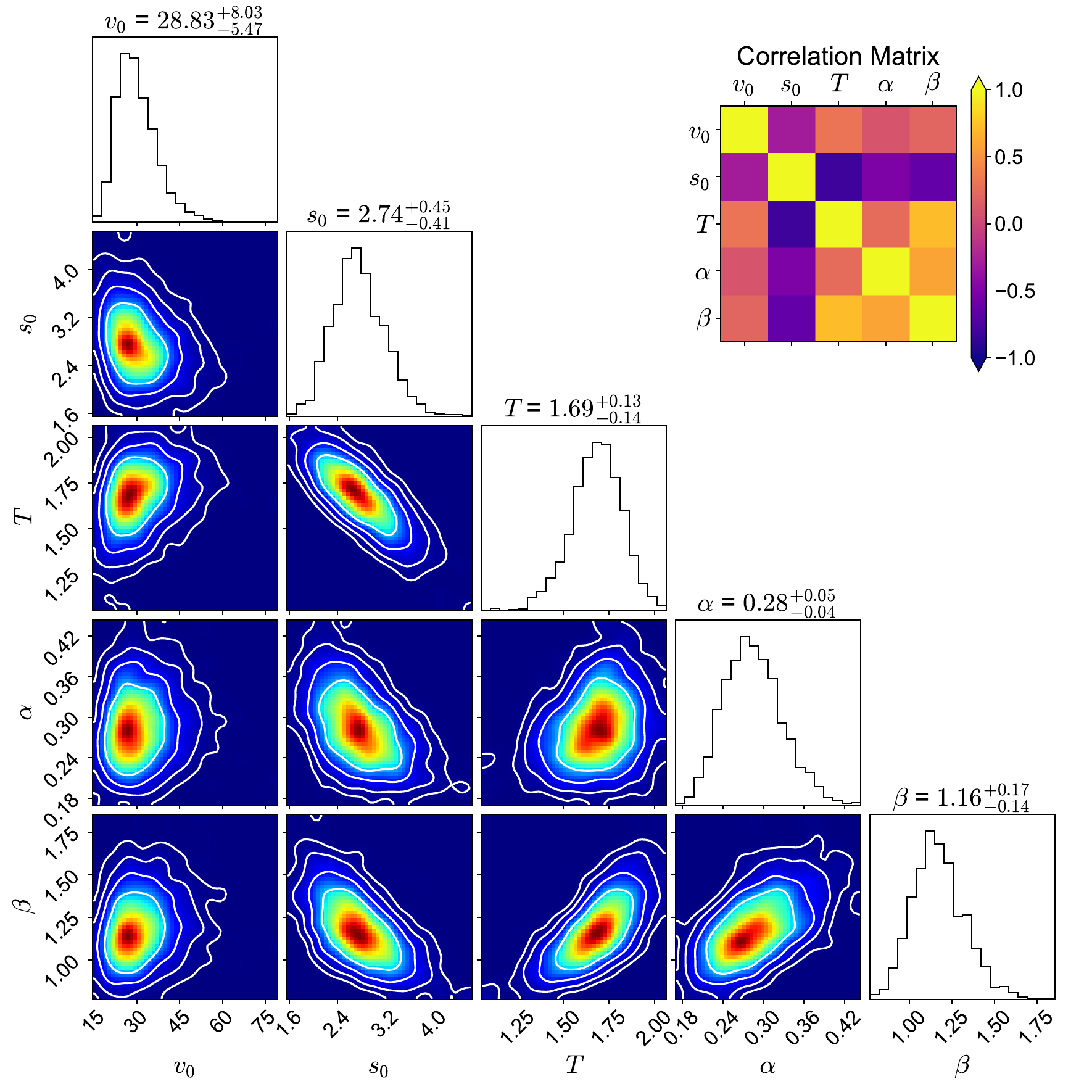}
    }
    \subfigure[Hierarchical MA-IDM posteriors of truck $\# 211$.]{
        \centering
        \includegraphics[width=0.483\textwidth]{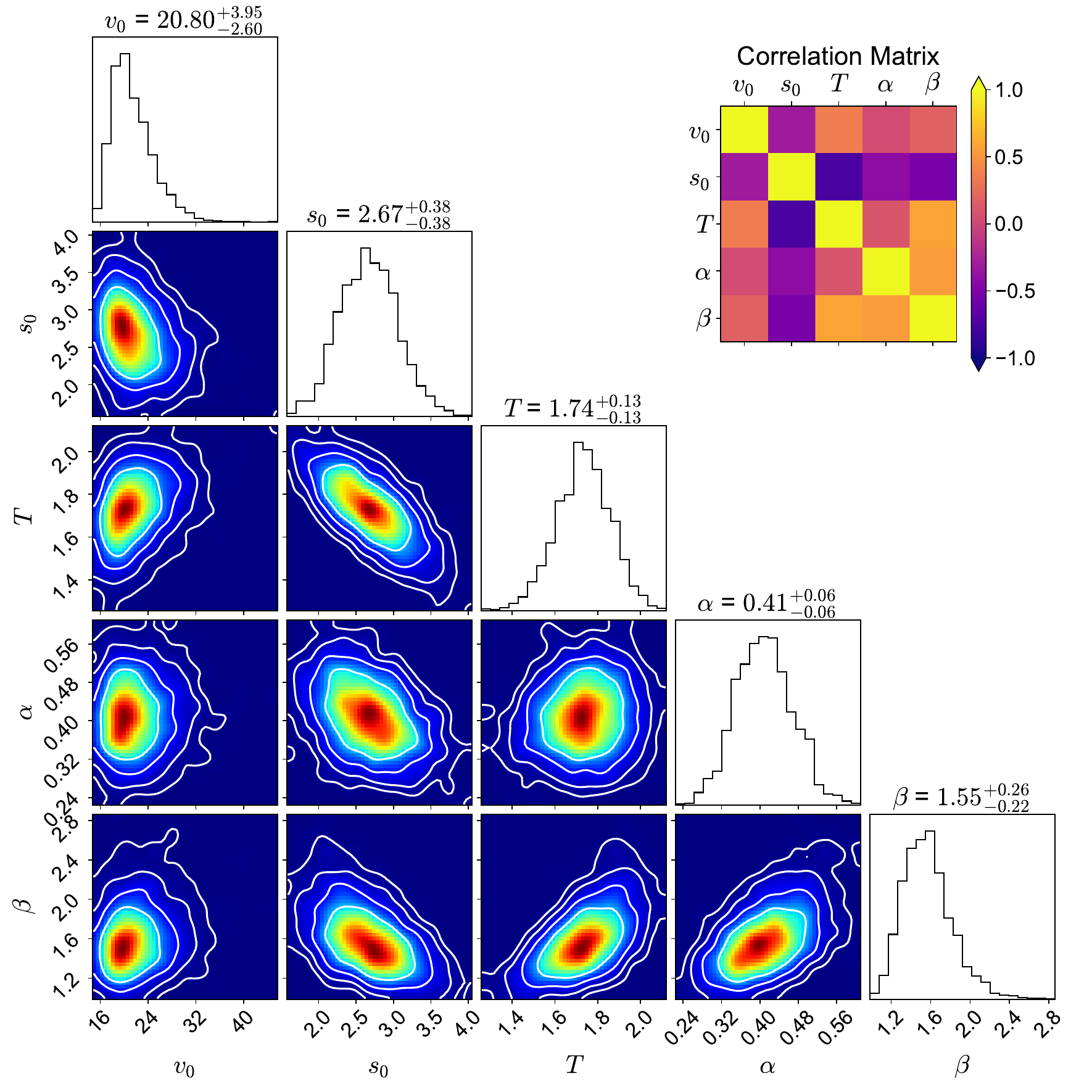}
    }
    \caption{The posterior distributions and the correlation matrices of the parameters of a truck driver $\#211$ in the hierarchical B-IDM and MA-IDM.}
    \label{fig:IDM_two_driver211}
\end{figure*}

\begin{table*}[!t]
\centering
\caption{Posterior Mean of Model Parameters.}
\label{posterior_expectations_table}
\begin{tabular}{ccccccc}
\toprule
Model & $\boldsymbol{\theta}=[v_0,s_0,T,\alpha,\beta]$  & $\sigma_\epsilon$ (m/s$^2$) & $\sigma_v$ (m/s) & $\sigma_x$ (m) & $\sigma_k$ (m/s$^2$) & $\ell\,(\mathrm{sec})$\\
\midrule
Pooled B-IDM (car)  & $[14.452,2.949,1.009,0.294,1.184]$ & $0.307$    & fixed& fixed& $\slash$& $\slash$\\
Pooled B-IDM (truck) & $[15.103,4.322,0.528,0.267,1.189]$  & $0.275$    & fixed & fixed& $\slash$& $\slash$\\
Hierarchical B-IDM ($\boldsymbol{\theta}$) & $[19.693,3.725,0.980,0515,1.529]$ & $0.204$  & fixed & fixed& $\slash$& $\slash$\\
Hierarchical B-IDM (car $\boldsymbol{\theta}_{\#273}$) & $[29.624,3.254,2.832,0.341,1.727]$ & $0.194$ & fixed & fixed& $\slash$& $\slash$\\
Hierarchical B-IDM (truck $\boldsymbol{\theta}_{\#211}$) & $[30.189,2.754,1.681,0.285,1.175]$ & $0.218$ & fixed & fixed& $\slash$& $\slash$\\
Unpooled B-IDM (car $\boldsymbol{\theta}_{\#273}$) & $[15.563,2.811,2.512,0.477,1.692]$ & $0.161$  & fixed & fixed& $\slash$& $\slash$\\
Unpooled B-IDM (truck $\boldsymbol{\theta}_{\#211}$) & $[16.837,1.862,1.796,0.589,1.852]$ & $0.178$ & fixed & fixed& $\slash$& $\slash$\\
\midrule
Pooled MA-IDM (car) & $[12.244,2.833,1.133,0.362,2.140]$   & $\slash$ & $0.005$ & $0.026$ & $0.390$       & $1.607$     \\
Pooled MA-IDM (truck)  & $[14.099,4.028,0.670,0.333,2.120]$  & $\slash$ & $0.005$ & $0.028$ & $0.354$       & $1.625$    \\
Hierarchical MA-IDM ($\boldsymbol{\theta}$) & $[16.919,3.538,1.183,0.553,2.147]$  & $\slash$  & $0.005$ & $0.027$ & $0.202$ & $1.435$\\
Hierarchical MA-IDM (car $\boldsymbol{\theta}_{\#273}$) & $[18.930,3.416,2.585,0.383,1.649]$   & $\slash$  & $0.005$ & $0.027$ & $0.195$ & $1.510$\\
Hierarchical MA-IDM (truck $\boldsymbol{\theta}_{\#211}$) & $[21.458,2.670,1.734,0.409,1.574]$    & $\slash$  & $0.005$ & $0.027$ & $0.214$ & $1.552$\\
Unpooled MA-IDM (car $\boldsymbol{\theta}_{\#273}$) & $[15.671,2.613,2.579,0.479,1.732]$ & $\slash$ & $0.004$ & $0.025$ & $0.172$       & $1.466$       \\
Unpooled MA-IDM (truck $\boldsymbol{\theta}_{\#211}$) & $[17.663,1.725,1.870,0.646,2.161]$  & $\slash$ & $0.004$ & $0.030$ & $0.183$       & $1.558$  \\
\bottomrule
\multicolumn{7}{l}{* Recommendation values \cite{treiber2000congested}: $\boldsymbol{\theta}_{\text{rec}}=[33.3, 2.0, 1.6, 1.5, 1.67]$.}

\end{tabular}
\end{table*}

\subsection{Calibration Results and Analysis}
Here we compare the calibration results of the pooled, hierarchical, and unpooled models for the B-IDM and the MA-IDM. The most straightforward results are the posteriors of the calibrated IDM parameters, for which we use \textit{corner.py} \cite{corner} to visualize. Here we adopt the hierarchical models as examples to illustrate our results. We visualized the posterior distributions of $\boldsymbol{\theta}_d$ (i.e., at the individual level) for two drivers in Fig.~\ref{fig:IDM_two_driver355} and Fig.~\ref{fig:IDM_two_driver211}, respectively. The posteriors of the B-IDM seem pretty similar to those of the MA-IDM. To make it clear to compare, we summarize the posterior mean of each parameter in Table~\ref{posterior_expectations_table}.

\subsubsection{Correlation matrices}
From the upper right part of each subfigure in Fig.~\ref{fig:IDM_two_driver355} and Fig.~\ref{fig:IDM_two_driver211}, the correlation matrices indicate several strong correlations among IDM parameters. Strong positive correlations exist in pairs of $(T,v_0)$, $(T,\beta)$, and $(\alpha,\beta)$; while strong negative correlations exist in the pairs of $(v_0, s_0)$, $(v_0,\alpha)$, $(s_0, T)$, $(s_0,\alpha)$, and $(s_0,\beta)$. These correlations are basically consistent with the results in the literature \cite{treiber2013microscopic} based on MLE. Also, the correlation matrices of $\boldsymbol{\theta}_d$ at the individual level are also consistent using hierarchical models and unpooled models.

\subsubsection{Analysis of the calibrated IDM parameters}\label{IDM_parameters_analysis}
We verify that all calibrated parameters fall into the recommendation ranges according to \cite{kesting2008calibrating}. The calibration results in Table~\ref{posterior_expectations_table} demonstrate the reliability of our methods to calibrate plausible IDM parameters. In the following, we make some comparisons among several pairs of calibration results.

Firstly, the results of the pooled model indicate the general driving behavior of a ``population driver,'' i.e., the model learns a set of IDM parameters with a moderate driving policy, which expects to explain all behaviors in the data. However, since we noticed that human driving behaviors vary from one driver to another due to disparate driving situations, the capability of the pooled model is limited in explaining heterogeneous data. Consequently, the pooled model considers diversity as errors to compensate for uncertain behaviors and thus assigns a large variance $\sigma_\epsilon^2$ to the error term. The unpooled model, on the other hand, is an example with small error variance $\sigma_\epsilon^2$, since it calibrates a separable model for each specific driver at the risk of overfitting.

Secondly, we obtain mostly similar identified parameters between the hierarchical and unpooled models at the individual driver's level. The results demonstrate the consistency of our proposed models with different hierarchies. Also, it strongly supports the claim that hierarchical models can simultaneously identify style-specific IDM parameters for diverse drivers.

\subsubsection{Analysis of kernel parameters for GP}\label{results_GP_analysis}
The experiment results are presented in Table~\ref{posterior_expectations_table} for the identified parameters of the MA-IDM. The SE kernel has two essential hyperparameters, the lengthscale $\ell_d$ and the kernel variance $\sigma_{k,d}^2$.

First, we focus on analyzing the lengthscale $\ell$. Theoretically, for $f\sim\mathcal{GP}(0, k)$, the correlation between $f(x)$ and $f(x')$ is $k(x,x)$, e.g., when $\|x-x'\|=\ell$, the correlation is $\exp(-1/2)\approx0.61$. Our experiments reveal that $\ell\approx1.4\sim1.6\,\mathrm{s}$ for the SE kernel. From the results, we conclude that in our case, the delay in the car follow-up action has a decent amount of positive correlation within $\ell $ (i.e., $\sim 1.6\,\mathrm{s}$); and is only slightly dependent within $\ell\sim3\ell$ (i.e., $1.6\sim 5\,\mathrm{s}$); but essentially no positive correlation outside of $3\ell$ (i.e., $\sim 5\,\mathrm{s}$). Additionally, both pooled and unpooled models show that truck drivers have longer reaction delays $\ell$ than car drivers.
We also noticed that the hierarchical and unpooled models have smaller $\ell$ than the pooled models, which is reasonable---the pooled models are less accurate and have to compensate for the uncertainty by increasing $\ell$ to obtain a higher likelihood.

Besides, experiments demonstrate that the white noise variance $\sigma_\epsilon^2$ of the B-IDM varies as we use disparate models or calibrate on different data. However, we would expect the random \textit{i.i.d.} noise to be at almost the same level for the same dataset since the observation noise is not supposed to vary substantially. Note that the variance terms in Eqs.~\eqref{eq_bidm_llh_x} and \eqref{eq_bidm_llh_x} are only identifiable as a whole, but the components cannot be separable. Therefore, we fix the observation noise $\sigma_x$ and $\sigma_v$. As for the MA-IDM, $\sigma_x$ and $\sigma_v$ stabilize at the same level for different models.

\subsection{Data Completeness and Parameter Orthogonality}\label{orthogonality}
Several questions are worth considering before calibrating the model: What kind of scenarios may the data contain to calibrate a good model? There are several parameters in the IDM formulation; are these parameters equally significant? Do we need to calibrate all the parameters \cite{punzo2014we}? The previous literature has reported that parts of the parameters could not be identified directly from the data due to insufficient information \cite{hoogendoorn2010calibration}. Researchers also developed a reduced IDM that fixes some unimportant parameters \cite{punzo2014we}. Furthermore, the identifiability of car-following systems is discussed in \cite{treiber2013traffic, wang2022identifiability}. In this part, we analyze these problems from the perspectives of data completeness and parameter orthogonality.

Data completeness and parameter orthogonality of IDM was first discussed by Treiber and Kesting \cite{treiber2013traffic, treiber2013microscopic}, in which the parameter orthogonality is illustrated as ``different identifiable driving situations are represented by disparate parameters, ideally one for each situation.'' We notice that in Eq.~\eqref{IDM_eq} the two parameters $v_0$ and $\delta$ are correlated with each other and orthogonal to the remaining parameters in $\boldsymbol{\theta}$. Specifically, they are shown in the same term $(v/v_0)^\delta$ but nowhere else. Given a certain $v$ and a specific quantity of this term, one can find infinite groups of the combination of $v_0$ and $\delta$. Therefore, when data is limited, experiments show that if we jointly calibrate $v_0$ and $\delta$, they would both converge to their priors. However, when data is sufficient to cover several approximate steady-state scenarios, the model is able to separately identify $v_0$ and $\delta$. Specifically, the combinations of $v_0$ and $\delta$ represent different unique dynamic characteristics of traffic flow---the fundamental diagram is triangular when $\delta\to+\infty$; while it is rounded when $\delta$ tends to be of low values \cite{treiber2013traffic}. Given that we only selected a limited number of drivers and for the sake of avoiding fractional exponential orders and numerically unstable issues, we follow the conventional settings and fix $\delta=4$ to obtain a parameter orthogonal model.

Next, we verify the completeness of the car-following data used in this work. The IDM parameters need to be learned from the corresponding data collected in specific scenarios. For instance, $v_0$ is learned from the free-flow data; $s_0$ and $T$ are learned from the steady-following data; $\alpha$ is learned from the freely accelerating data, and $\beta$ is learned from the approaching (with braking) data. However, since we only selected the interactive stop-and-go data, the maximum speed in the data was never higher and there are no free traffic regimes, so the desired speed $v_0$ is hard to identify here. The experiments are consistent with our expectation---from Table~\ref{posterior_expectations_table}, we can see that the identified $v_0$ varies in a large range for disparate models. The Bayesian approach is particularly adept at addressing these nuances. Our experiments reinforce the idea that the posterior's dependency on the prior is a function of its strength and the adequacy of the data. For instance, a noninformative prior results in a non-identifiable $v_0$, while a strong prior closely aligns $v_0$ with its set value. These findings provide a practical demonstration of the fundamental principles of Bayesian analysis in model calibration. Here we emphasize the Bayesian principle that the posterior heavily leans on the data in the case of weak (or noninformative) priors, whereas it tends to mirror the prior if the prior is strong (low variance) or if the data is insufficient. Additionally, when tuning the values of the priors, the correlations between parameters could result in noticeable changes in other posteriors, illustrating the interdependent nature of these variables in the Bayesian framework. This analysis serves as a valuable example of the dynamics between data sufficiency and prior influence in Bayesian investigations.

\section{Simulation}\label{simulation_sec}

\begin{figure*}[t]
    \centering
    \subfigure[Simulated motion states with hierarchical models.]{
        \centering
        \includegraphics[width=0.486\textwidth]{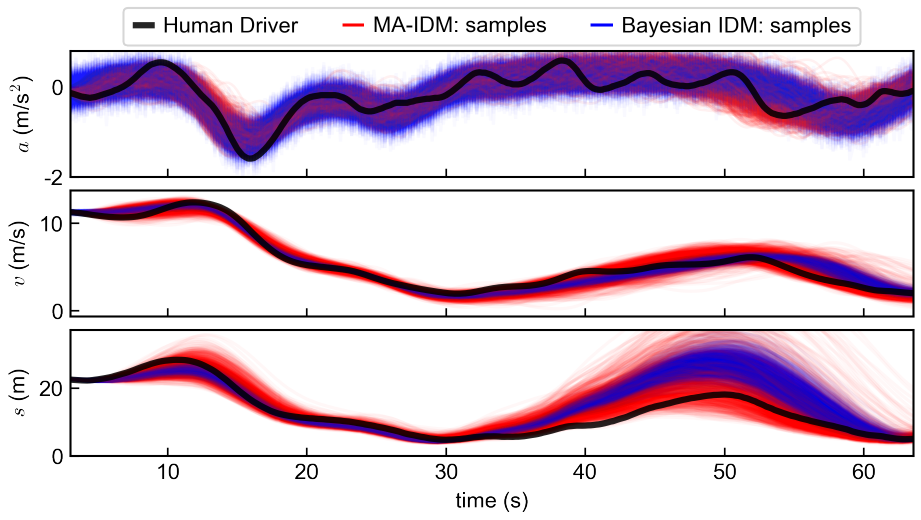}
    }\hspace{3.7mm}
    \subfigure[Simulated motion states with unpooled models.]{
        \centering
        \includegraphics[width=0.46\textwidth]{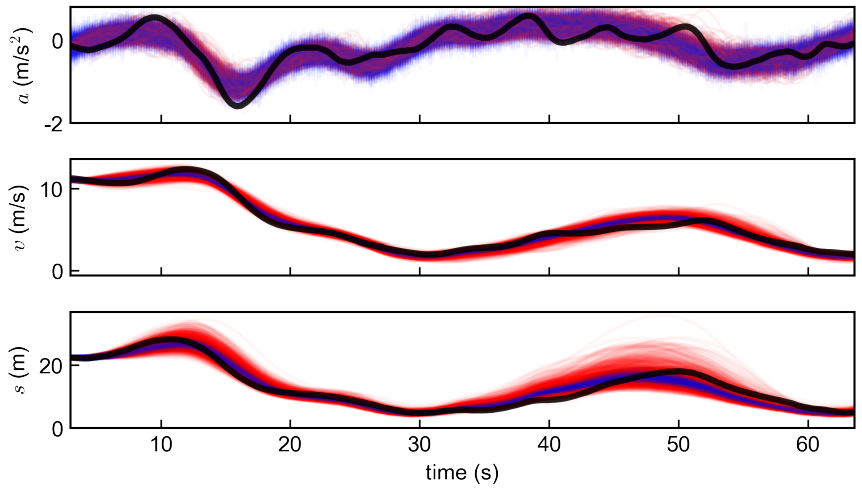}
    }
    \caption{The stochastic simulation results of a representative truck driver $\# 211$. The black lines indicate the ground truth of human driving data. The red and blue lines are the simulated motion states with the parameter samples drawn from posteriors.}
    \label{fig:simulation_for_two_cars}
\end{figure*}

\begin{table*}[t]
    \centering
    \caption{Evaluations of the short-term (3 s) simulations with different proposed models.}
    \begin{tabular}{c|c|c|c|c|c|c|c}
    \toprule
        \multicolumn{2}{c|}{\multirow{3}{*}{ \begin{tabular}{c}
            Data $\backslash$ Model
        \end{tabular} }} & \multicolumn{2}{c|}{Pooled model} & \multicolumn{2}{c|}{Hierarchical model} & \multicolumn{2}{c}{Unpooled model} \\
    \cmidrule(lr{.2em}){3-8}
        \multicolumn{2}{c|}{} & {B-IDM}  & {MA-IDM} & B-IDM & {MA-IDM} & {B-IDM} & {MA-IDM} \\
    \midrule
        \multirow{6}{1cm}{\centering Car $\#273$} & $e(a) $ (m/s$^2)$ & $3.87\pm0.89$ & $2.16\pm1.47$ & $2.78\pm0.56$ & $1.18\pm0.77$ & $2.17\pm0.57$ & {$\mathbf{1.04\pm0.67}$} \\
         & $e(v)$ (m/s) & $3.95\pm2.32$ & $2.27\pm1.60$ & $2.39\pm1.34$ & $1.30\pm0.92$  & $2.14\pm1.49$ & $\mathbf{1.15\pm0.83}$ \\
         & $e(s)$ (m) & $4.66\pm2.52$ & $2.42\pm1.60$ & $3.46\pm1.96$ & $\mathbf{2.24\pm1.59}$ & $3.52\pm2.14$ & $\mathbf{2.24\pm1.59}$ \\
         \cmidrule(lr{.2em}){2-8}
         & ${\mathrm{CRPS}}(a)$ & $1.46\pm0.55$ & $1.09\pm0.81$ & $0.96\pm0.28$& $0.58\pm0.38$  & $0.89\pm0.44$ & $\mathbf{0.53\pm0.39}$ \\
         & ${\mathrm{CRPS}}(v)$ & $2.65\pm1.89$ & $1.06\pm0.81$ & $1.50\pm0.93$ & $0.63\pm0.47$ & $1.47\pm1.22$ & $\mathbf{0.56\pm0.43}$ \\
         & ${\mathrm{CRPS}}(s)$ & $3.05\pm1.71$ & $1.87\pm1.46$ & $2.41\pm1.60$ & $\mathbf{1.83\pm1.49}$ & $2.65\pm1.83$ & $1.84\pm1.50$ \\
        \midrule
        \multirow{6}{1cm}{\centering Truck $\# 211$} & $e(a)$ (m/s$^2$) & $3.66\pm0.79$ & $2.23\pm1.42$  &$2.74\pm0.62$ & $\mathbf{1.15\pm0.63}$ & $2.58\pm0.61$ & ${1.17\pm0.75}$ \\
         & $e(v)$ (m/s) & $3.75\pm1.94$ & $2.53\pm1.84$ & $2.32\pm1.21$ & $\mathbf{1.39\pm0.91}$ & $2.24\pm1.03$ & ${1.52\pm1.12}$\\
         & $e(s)$ (m) & $4.64\pm2.30$ & $2.84\pm1.58$ & $3.84\pm2.74$ & $\mathbf{2.46\pm1.62}$ & $3.77\pm2.71$ & ${2.56\pm1.66}$\\
        \cmidrule(lr{.2em}){2-8}
         & ${\mathrm{CRPS}}(a)$ & $1.44\pm0.44$ & $1.31\pm1.02$ & $1.11\pm0.39$ & $\mathbf{0.61\pm0.34}$ & $1.11\pm0.42$ & ${0.74\pm0.55}$ \\
         & ${\mathrm{CRPS}}(v)$ & $2.69\pm1.68$ & $1.40\pm1.17$  & $1.62\pm0.97$ & $\mathbf{0.73\pm0.51}$ & $1.63\pm0.84$ & ${0.93\pm0.75}$\\
         & ${\mathrm{CRPS}}(s)$ & $3.14\pm1.65$ & $2.16\pm1.38$ & $2.88\pm2.42$ & $\mathbf{1.96\pm1.48}$ & $2.91\pm2.44$ & $2.09\pm1.52$\\
    \bottomrule
    \multicolumn{8}{l}{* All values in the table are amplified ten times to keep an efficient form.}
    \end{tabular}
    \label{tab:errors}
\end{table*}

Deterministic simulation is a commonly used approach in trajectory calibrations \cite{treiber2013microscopic}. Typically, given the leading vehicle's trajectory and the initial motion state of the following vehicle, we can use the calibrated IDM model to conduct simulations and obtain predicted motion states; then compare the predicted motion states with the observations to further help with either calibration or evaluation. In the previous literature, most motion state updating methods are based on Eq.~\eqref{simulation_update}, which is deterministic. However, such deterministic simulations limit the simulator's ability to reproduce human-like driving behaviors. Furthermore, as we clarified in Section~\ref{unbiased_model}, only biased models would be obtained without considering serial correlation, and the deterministic simulations are no exception.

Benefiting from the Bayesian method, we can obtain posterior motion states in stochastic simulations rather than the deterministic simulations in the literature. In the following, we illustrate the proposed stochastic simulation methods with both short-term and long-term simulations. The short-term simulation directly reflects how well the model is calibrated using the data, thus reflecting the regression performance. The long-term simulation focuses more on how well it captures human driving behaviors and traffic dynamics, thus verifying the ability of the well-calibrated car-following models. Then, the quantitative analysis is demonstrated for comparison.

\subsection{Parameters Generation and Stochastic Simulations}
It was reported in the previous literature that car-following parameters simply drawn from uncorrelated marginal distributions could yield unreliable results in simulation and, consequently, inaccurate interpretation \cite{kim2011correlated}. Here we emphasize that one advantage of our Bayesian method compared with the traditional optimization-based solution is obtaining the joint distributions of parameters instead of point estimations. Therefore, rather than using the mean values from Table~\ref{posterior_expectations_table}, one can draw enormous samples from the joint posterior distributions (e.g., Fig.~\ref{fig:IDM_two_driver355}) to generate human-like driving behaviors. Specifically, from the viewpoint of a generative process, it is not difficult to sample the parameters of the distribution of $\boldsymbol{\theta}_d, d=1,\dots,D$, from the posteriors of $\boldsymbol{\theta}$ (at the population level) to obtain disparate individual driving styles; and sample style-specific IDM parameters $\boldsymbol{\theta}_{d,i}, i=1,\dots,N$ from the posteriors of $\boldsymbol{\theta}_d$ (at the individual level) to obtain enormous but nonidentical driving behaviors.

Given that the intrinsic nature of our Bayesian model is probabilistic distribution, here we propose a memory-augmented method by considering the serial correlation in stochastic simulations to obtain unbiased posterior motion states. The key idea is to sample an action $a_d^{(t)}$ instead of using a deterministic $a_{\mathrm{IDM},d}^{(t)}$. The stochasticity of such simulation is determined by the noise level $\sigma_\epsilon$ or $\sigma_k$, calibrated from the data as discussed in Section~\ref{results_GP_analysis}. Specifically, recall that we assumed $a_d^{(t)} = a_{\mathrm{IDM},d}^{(t)} + a_{\mathrm{GP},d}^{(t)}$ in Section~\ref{unbiased_model}, we simulate the motion of the following vehicle at a certain timestamp by simultaneously taking several independent steps as follows. To initialize, we draw several random functions $a_{\mathrm{GP},d}$ for $d=1,\dots,D$ from GP at the beginning of the simulations. Then for each driver $d$ at timestamp $t$, we separately draw samples for the two terms in Eq.~\eqref{MA_idm_eq}:
\begin{enumerate}
    \item Obtain the first term $a_{\mathrm{IDM},d}^{(t)}$ by feeding $\boldsymbol{\theta}_d$ and inputs into the IDM function;
    \item Draw a sample $a_{\mathrm{GP},d}^{(t)}|\boldsymbol{a}_{\mathrm{GP},d}^{(t-T:t-1)}$ at time $t$ from the GP to obtain the temporally correlated information $a_{\mathrm{GP},d}^{(t)}$;
\end{enumerate}
Repeat these two steps until a certain simulation time. To efficiently sample from step 2, one can adopt an efficient computational simulation method for GP \cite{doucet2010note}, which bypasses the computation of posterior covariance in GP. Without loss of generality, let us denote the past action residuals as $\boldsymbol{\hat{y}}=\boldsymbol{a}_{\mathrm{GP},d}^{(t-T:t-1)}$ and the future action residuals as ${\hat{x}}={a}_{\mathrm{GP},d}^{(t)}$. Two simple but essential steps construct the efficient method: (i) sample a random vector
\begin{equation}
\left(\begin{array}{c} x\\\boldsymbol{y}\end{array}\right)\sim \mathcal{N}\left(\boldsymbol{0},
    \boldsymbol{K}=\left[\begin{array}{cc}
        \boldsymbol{\Sigma}_{xx} & \boldsymbol{\Sigma}_{xy} \\
        \boldsymbol{\Sigma}_{yx} & \boldsymbol{\Sigma}_{yy}
    \end{array}\right]\right),
\end{equation}
and (ii) compute $x+\boldsymbol{\Sigma}_{xy}\boldsymbol{\Sigma}_{yy}^{-1}(\boldsymbol{\hat{y}}-\boldsymbol{y})$ as a sample for $a_{\mathrm{GP},d}^{(t)}|\boldsymbol{a}_{\mathrm{GP},d}^{(t-T:t-1)}$. For (i), we can precompute and save the Cholesky factor $\boldsymbol{L}$ that gives $\boldsymbol{\Sigma}=\boldsymbol{L}\boldsymbol{L}^{\top}$, and therefore efficient sampling can be achieved by first generating a standard Gaussian sample $\boldsymbol{u}\sim \mathcal{N}(\boldsymbol{0},\boldsymbol{I})$ and then performing a transformation $\boldsymbol{L}\boldsymbol{u}$. For (ii), once again we can precompute the matrix $\boldsymbol{\Sigma}_{xy}\boldsymbol{\Sigma}_{yy}^{-1}$ and use it to compute all conditional samples. With this, the stochastic simulation only involves sampling from standard Gaussian distribution and performing matrix-vector multiplication, so the overall simulation process is very efficient.

\subsection{Error Analysis and Bayesian Evaluations}\label{stoch_sim}
To evaluate the calibration results, we use one representative driver for each vehicle type, i.e., the car driver $\# 273$ and the truck driver $\# 211$. We fix the IDM parameters to their mean values for each model. Then, by providing the leading vehicle's trajectories and starting the followers with the same initial conditions, we simulate the trajectories of the followers, whose behaviors are controlled by a calibrated IDM with a set of sampled parameters \cite{ciuffo2008comparison}. The simulation step is set as $0.2$ s which corresponds to the $5$ Hz sampling frequency.

To illustrate the simulation results, we first show the long-term simulated trajectories generated by the IDM parameters sampled from $\boldsymbol{\theta}_{\#211}$ in Fig.~\ref{fig:simulation_for_two_cars}. It demonstrated that the calibrated car-following models (both B-IDM and MA-IDM) can capture the human driver's behavior---accelerate to follow the leader and brake to avoid collision.

\begin{figure}[t]
    \centering
    \includegraphics[width=\linewidth]{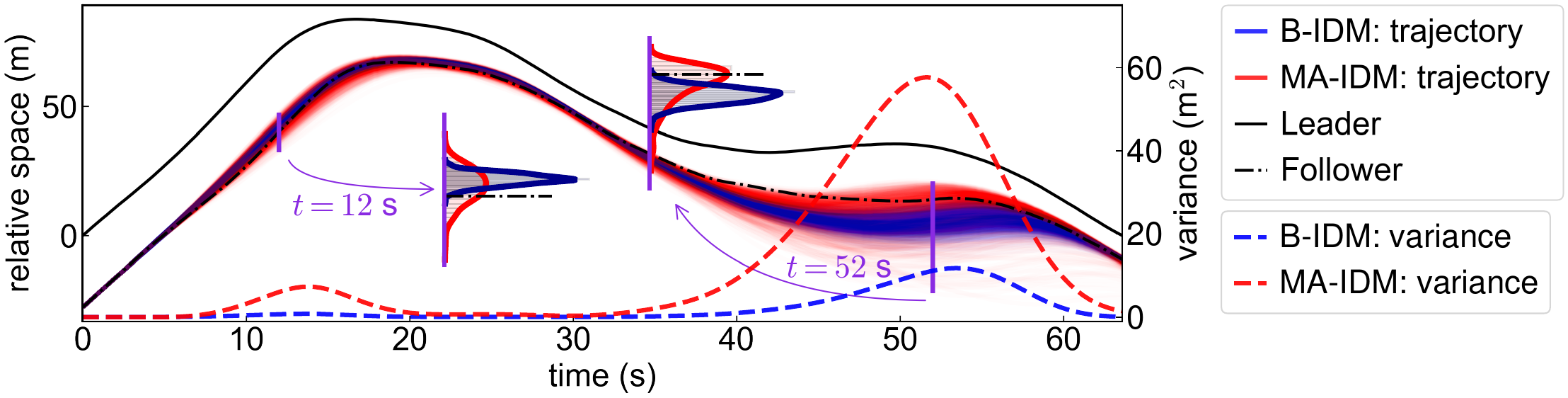}
    \caption{The time-space diagram of the follower's posterior trajectories with hierarchical models at the point of view of another `observing' vehicle with a constant mean speed.}
    \label{fig:CFpairs}
\end{figure}

In the following, we evaluate the regression performances by comparing the short-term ($3$ s) simulated trajectories with the ground truth via measuring the root mean square errors (RMSE) $e$ on their motion states (i.e., $a$, $v$, and $s$), written as
\begin{align}
    e(x, \boldsymbol{\hat{f}},\boldsymbol{\theta}_{d,i}) =\sqrt{ \frac{1}{N} \sum_{i=1}^N \left(f(x_i;\boldsymbol{\theta}_{d,i})-\hat{f}_i\right)^2}.
\end{align}

Short-term simulation errors are reported in Table~\ref{tab:errors} in the form of (mean $\pm$ standard deviation), where the models with the lowest error are in bold font. Recall that the pooled models, acting as general behavioral models, were trained to fit all the car-following data of several drivers. Apparently, given that different drivers have distinct driving styles and have experienced different driving situations, the data could reflect diverse operating characteristics. Thus, a model that best fits a particular driver does not necessarily do so for a different driver \cite{kesting2008calibrating}. Therefore, it is reasonable that the pooled models have limited performances. Besides, the simulation results indicate that the errors of the hierarchical models are at the same level as the unpooled models, which agrees with the findings in Section \ref{IDM_parameters_analysis} that the hierarchical models can identify disparate IDM parameters of diverse drivers. As shown in Fig.~\ref{fig:CFpairs}, the predicted motion states with the calibrated MA-IDM tend to be closer to ground truth when compared with those driven by the calibrated B-IDM.

To evaluate the performance of stochastic simulations and quantify the uncertainty of posterior motion states, we adopted the continuous ranked probability score (CRPS) \cite{matheson1976scoring}, which can be written as
\begin{equation}
    \mathrm{CRPS}(y_t) = \int_{-\infty}^{+\infty}\left(F(y)-\mathbbm{1}\{y>y_t\}\right)^2 dy,
\end{equation}
where $y_t$ is the observation at time $t$, $F$ is the forecast cumulative distribution function, and $\mathbbm{1}$ is the indicator function. To build the evaluation metric, we first evaluate CRPS for $a$, $v$, and $s$ at each timestamp $t$, and then average the values over time by $\overline{\mathrm{CRPS}}(y)= \frac{1}{T}\sum_{t=1}^T \mathrm{CRPS}(y_t)$.

From Fig.~\ref{fig:simulation_for_two_cars} and Fig.~\ref{fig:CFpairs}, an interesting but not surprising finding is that the red lines tightly contain the ground truth in the envelope. In addition, although the B-IDM (the blue lines) loosely and easily contains the ground truth of the acceleration curve in its envelope, the simulated results for $v$ and $s$ still deviate from the ground truth. From Table~\ref{tab:errors}, we conclude that by introducing the memory-augmented method to model the serial correlation, the MA-IDM with unbiased parameters outperforms the biased B-IDM.

\begin{figure}[t]
    \centering
    \subfigure[Simulation with fixed IDM parameters and random white noise.]{
        \centering  \includegraphics[width=.48\textwidth]{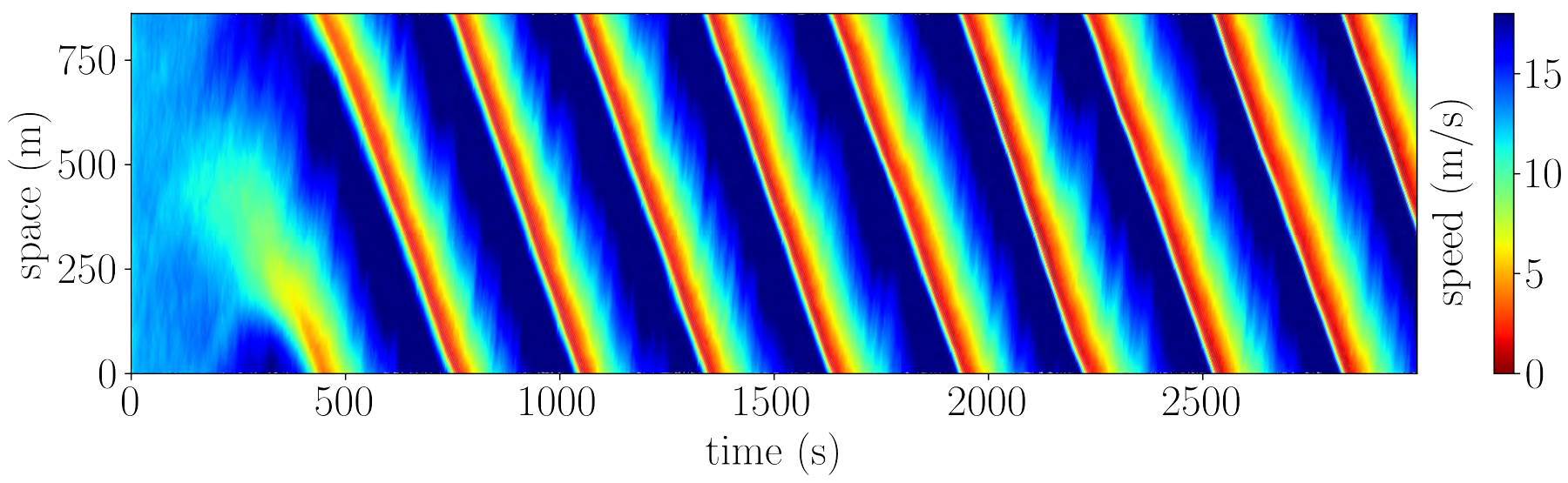}
    }
    \subfigure[Simulation with MA-IDM and GP.]{
        \centering\includegraphics[width=.48\textwidth]{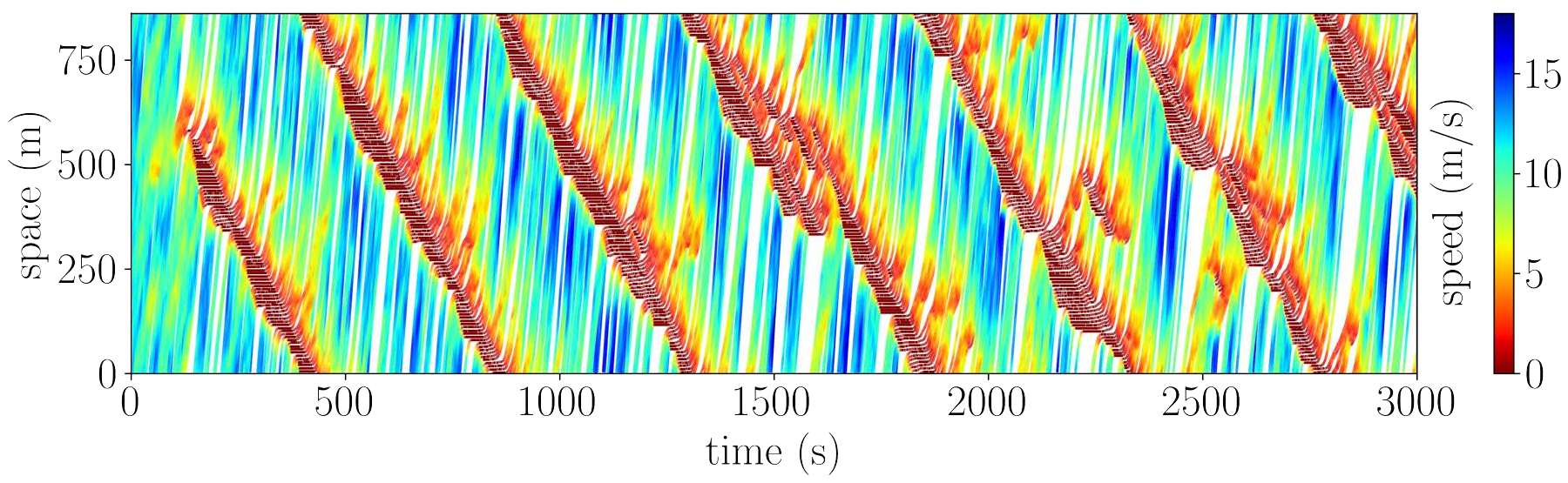}
    }
    \caption{Time-space diagram of multi-vehicle car-following simulations. (a) the parameters are taken as the recommendation values \cite{treiber2000congested}. (b) the parameters are sampled from the posteriors in the hierarchical MA-IDM.}
    \label{fig:multi_vehicle_simulation}
\end{figure}

\subsection{Multi-Vehicle Simulations in A Ring Road Scenario}
The single-lane ring road network has been extensively studied to investigate microscopic car-following behaviors in a closed traffic system \cite{treiber2010open, sugiyama2008traffic}. Several key elements determine the dynamics on the ring road: The radius is set at $128$ m; The initial speed of each vehicle is set at $11.6$ m/s; There are a total of $37$ vehicles simulated for $3000$ sec; The simulation step is set at $0.2$ sec.

In Fig.~\ref{fig:multi_vehicle_simulation}, we perform simulations with two different settings to compare microscopic behaviors. Figure~\ref{fig:multi_vehicle_simulation} (a) is simulated with homogeneous IDM parameters taken as the recommended values listed in Table~\ref{posterior_expectations_table}. Figure~\ref{fig:multi_vehicle_simulation} (b) shows the results of the simulations, in which the GP settings are the same as in Section~\ref{stoch_sim}. As can be seen, we can observe a recurring pattern from the simulations with fixed IDM parameters, although a stochastic term is introduced. On the contrary, we can obtain different random car-following behaviors in the heterogeneous setting, which reflects the dynamic and diverse driving styles of different drivers.

\section{Conclusion and Future Scope}\label{conclusion}
This paper takes IDM as an example and develops a novel calibration method (i.e., the MA-IDM with three hierarchies) by leveraging Bayesian methods and Gaussian processes, which jointly identifies the IDM parameters and models the autocorrelated errors, thus obtaining optimal parameter estimation. We conducted extensive experiments to compare the pooled, hierarchical, and unpooled models for both B-IDM and MA-IDM. Experiment results demonstrate the effectiveness of the MA-IDM and reveal that taking the past 5-second driving actions into account can be helpful in modeling and simulating the human driver's car-following behaviors. Lastly, with the well-calibrated MA-IDM, a novel stochastic simulator is proposed to perform stochastic simulation efficiently.

Several aspects can be explored further in future work. First, our method is not limited to calibrating IDM; one can easily extend it to other microscopic car-following models (even other regression problems, see, e.g., \cite{cheng2022bayesian}) using a similar Bayesian scheme introduced in this work. In addition, the GP component is not restricted to only using the SE kernel. One can easily make minor adjustments to use other kernels, e.g., Mat\'{e}rn kernel. Note that the Ornstein-Uhlenbeck (OU) process, another well-known stochastic process with memory characteristics, could also be integrated into our analytical framework. The OU process shares significant similarities with order-1 autoregressive (AR) processes with a positive coefficient and Gaussian processes (GPs) employing a Mat\'{e}rn 1/2 kernel. Additionally, human drivers have personalized styles, which are not life-long determined but time-varying; thus, it is also interesting to explore: ``how to model the dynamic time-varying IDM parameters for every single driver? \cite{zhang2022generative}'' Following this concern, we further emphasize that, in reality, the error variance and lengthscale should be scenario-specific. For instance, one could freely take actions with a significant variance in free flow, but most drivers take almost the same actions (e.g., hard brake) in some risky scenarios. Thus, the GP hyperparameters (e.g., noise variance and temporal lengthscale) should be input-dependent, and the GP itself will become nonstationary. Furthermore, more realistic scenarios involving multiple factors can be studied by extending our model into a more complex architecture. For instance, human driving behaviors result from frequent transitions among various driving styles \cite{zhang2021spatiotemporal}. Thus, it is no longer reasonable to model driving behaviors using a single distribution; instead, a mixture model is suitable \cite{wang2018learning}. Furthermore, as mentioned in \cite{Michail2022}, the power dynamics of vehicles is not uniform throughout the speed range; also, different vehicles (e.g., cars and trucks in this work) have differences in power capabilities. These factors lead to heterogeneity in vehicle dynamics \cite{ciuffo2018capability, Michail2022}.

{\normalem
\bibliographystyle{IEEEtran}
\bibliography{references}}

\end{document}